\documentclass[useAM,use natbib]{mn2e}

\usepackage{psfig, epsf, epsfig}
\usepackage{caption}
\usepackage{graphicx}
\usepackage{float}
\usepackage{placeins}
\usepackage{epstopdf}

\title[Dark matter masses in galaxies]{Estimating galaxy masses
from kinematics of globular cluster systems:
a new method based on deep learning} 
\author[Rajvir Kaur, Kenji Bekki, Ghulam Mubashar Hassan and Amitava Datta]
{Rajvir Kaur${}^1$\thanks{corresponding email:
22397217@student.uwa.edu.au},
Kenji Bekki${}^2$,
Ghulam Mubashar Hassan${}^1$
and Amitava Datta${}^1$ \\ 
${}^1$
Department of Computer Science and Software Engineering, 
The University of Western Australia, Australia\\
${}^2$ International Centre for Radio Astronomy Research (ICRAR), The University of Western Australia, Australia \\
}

\begin{document}

%\date{Accepted, Received 2005 February 20; in original form }

\pagerange{\pageref{firstpage}--\pageref{lastpage}} \pubyear{2005}

\maketitle

\label{firstpage}

\begin{abstract}

We present a new method by which the total masses of galaxies including
dark matter can be estimated from the kinematics of 
their globular cluster systems (GCSs). 
In the proposed method, we apply the convolutional neural networks (CNNs) 
to the two-dimensional (2D) maps of line-of-sight-velocities ($V$)
and velocity dispersions ($\sigma$) of GCSs predicted from numerical simulations
of  disk and elliptical galaxies.
In this method,  we first train the CNN using either only 
a larger number ($\sim 200,000$) of the synthesized 2D maps of $\sigma$
(``one-channel'') or those of both $\sigma$ and $V$ (``two-channel'').
Then we use the CNN to predict the total masses of galaxies (i.e., test the CNN)
for the totally unknown dataset that is not used in training the CNN. 
The principal results show that overall accuracy for  one-channel
and two-channel data is 97.6\% and 97.8\% respectively,
which suggests that the new method is promising.
The mean absolute errors (MAEs) 
for one-channel and two-channel data are 0.288 and 0.275 
respectively, and the value of root mean square errors (RMSEs) 
are 0.539 and 0.51 for one-channel and two-channel respectively. 
These smaller MAEs and RMSEs 
for two-channel data
(i.e., better performance)
suggest that the new method can properly consider
the global rotation of GCSs in the mass estimation.
We also applied our proposed method to real data collected from observations of NGC 3115 to compare
the total mass predicted by our proposed method and other popular methods from the literature.

\end{abstract}


\section{Introduction}

Globular clusters systems (GCSs) of galaxies have long been considered to
have fossil information not only on the physical properties of their host
galaxies but also on the their hosts' formation histories
\citep[e.g.,][]{Brodie2006, Forbes2018}.
Therefore, the physical properties of GCSs such as  the projected
radial number distributions,  kinematics,  metallicity distribution functions,
and numbers of globular clusters (GCs) per unit galaxy luminosity 
(``specific frequency'')
have been investigated by both observational and theoretical astronomers 
in various contexts of galaxy formation. For example, the radial density
profiles of GCSs have been discussed in the context of suppression of
star formation (thus GC formation) in dwarf galaxies by cosmic reionization
effects, \citep[e.g.,][]{Santos2003, Bekki2008, Spitler2012, Griffen2013, Boylan-Kolchin2018}

One of the great advantages in studying GCSs in the context of galactic properties is
that their kinematics can be used to estimate the total masses of galaxies including
dark matter halos well beyond the sizes of
their stellar distributions, \citep[e.g.,][]{Romanowsky2003, Peng2004, Alabi2016}. Accordingly, the previous  mass estimation of galaxies well beyond
their five  effective radii ($R_{\rm e}$) have provided strong constants on the
total masses  of dark matter and their mass fractions, 
both of which can be compared with corresponding theoretical predictions,
\citep[e.g.,][]{Alabi2017}. Recently, the mass estimation of dark matter in
ultra-diffuse galaxies (UDGs) using the kinematics of GCSs made a significant
contribution to the better understanding of the UDG formation  in clusters
of galaxies \citep[e.g.,][]{Beasley2016}. The mass estimation of galaxies using
GCS kinematics is considered to be complementary to other mass estimation
methods  using different  mass tracers
such as planetary nebulae (PNe), \citep[e.g.,][]{Peng2004,Morganti2013}
and X-ray gas, \citep[e.g.,][]{Su2014}.

There are a number of ways to estimate the total 
masses of galaxies using the radial number density profiles and kinematics
of mass tracers like GCs and PNe  in galaxies \citep[e.g.,][]{Alabi2016}.
For example,
\citep{Romanowsky2003} used the observed line-of-sight-velocities ($V$)
of 109 PNe in NGC 3379 to estimate the total mass  and the mass-to-light-ratio
using the spherical
Jeans equation with a range of the anisotropy parameter.
\citep{Peng2004}  used the radial profiles of GC number density,
velocity dispersion, and rotation amplitude
in NGC 5128 in order to derive the total mass based on
the Jeans equation. 
More recently, \citep{Alabi2016} 
have
applied the new robust mass estimator proposed by  Watkins et al (2010)
to the observed kinematics of GCSs in 23 early-type galaxies from
SLUGGS survey \citep{Brodie2014} in order to derive the total masses
of galaxies and the mass fractions of dark matter up to  $\sim 13R_{\rm e}$.

Deep learning has been recently used for various purposes of astronomical studies,
such as morphological classification of galaxies, \citep[e.g.,][]{Dieleman2015}
and determination of mass-ratios of galaxy merging leading to S0 formation,
\citep[e.g.,][]{Diaz2019}. 
These new  methods based on deep learning can dramatically speed up
astronomical tasks (e.g., automated galaxy classification without using human eye) and thus improve the productivity in astronomical research.
One of key requirements in deep learning based techniques is to use a large ``training dataset'' to train the deep learning networks.
It is thus possible that deep learning can be applied to the mass estimation
of galaxies using kinematics of GCs, 
if a large number of
two-dimensional (2D) maps of GCS kinematics are available for training the deep learning network.

A possible advantage of the mass estimation method based on deep learning
is that the details of 2D kinematics of GCSs can be all used to derive
the total masses of galaxies: radial and azimuthal variations of 
line-of-sight-velocities ($V$, which can measure global rotation)
and velocity dispersions ($\sigma$, which can measure
random kinetic energy) can be used for the mass estimation in the method
based on deep learning.
Although the radial profiles
of velocity dispersions ($\sigma$) of GCSs can be self-consistently
considered in  the standard method based on the Jeans equation,
the rotation amplitudes of GCSs and their radial and azimuthal dependencies
are not properly included in the method.
Also, the tracer mass estimator of \citep{Watkins2010} assumes
spherically symmetric distributions of GCs and thus does not consider properly
global rotation (along minor and major axes) of GCSs  and its radial dependence.
These suggest that the new mass estimation method based on deep learning
can be promising, if a large number of 2D kinematics of GCSs can be used.

The purpose of this paper is thus to estimate the total mass of
disk and elliptical  galaxies by applying
deep learning to the kinematics of GCSs of the galaxies.
Since this is the very first paper on this topic,
we try to demonstrate that the new method does work in the mass estimation
using the synthesized two-dimensional (2D) maps of line-of-sight velocities
($V$) and velocity dispersion ($\sigma$) from numerical simulations
of these galaxies with GCs. 

%\textbf{ We have also investigated the observational data of the globular cluster system (GCS) in NGC 3115 \citep{Dolfi2020} as one of the authors in this article is on \citep{Dolfi2020} which facilitated the access to the data. In our experiments with observational data, we have found that the GCs in the galaxy are not distributed across the square field. Therefore, we cannot have the reliable 2D map of GCS kinematics for the square region around the galaxy similar to simulated square maps of GCSs. Thus, this is not an ideal situation as we need the square map of GCS kinematics for the input to proposed CNN model. However, we estimated the missing data in 2D map of GCS kinematics to make a square region around galaxy in different ways. We realise that there should be other datasets for other galaxies, but it is beyond the scope of this paper to generate the square 2D maps of GCS kinematics. Therefore, we would like to handle this problem in our future studies.}

In this new method based on deep learning,
 we first need to generate a large number of pairs of 
(i) 2D maps of GCS kinematics and (ii) known total masses of galaxies (``labels'')
from numerical simulations in order to train the proposed convolutional neutral network (CNN). 
In this ``supervised learning'', a large number of synthesized images
of GCSs ($>10^4$) are used to train the proposed model.

We mainly investigate how accurately the CNNs can predict the total masses
of galaxies whose 2D maps of GCSs are not used in the training phase 
of CNNs. We also investigate how the prediction accuracy can depend on
the introduction of noise in the synthesized 2D maps of GCSs, because
it is inevitable for spectroscopic estimation of $V$ for GCs 
to have a certain degree of observational errors.
We use the 2D kinematics maps of GCSs in isolated disk galaxies and 
elliptical galaxies formed from
major merging of two disk galaxies  with different orbital parameters.
We adopt such  idealized major merger models for elliptical galaxy formation,
because this study is still in the proof-of-concept phase for the adopted
new mass estimation method. We will use the 2D maps of GCSs from cosmological
simulations of GCS formation in our future studies.

The plan of the paper is as follows.
We describe (i) the models of disk and elliptical galaxies with GCS,
(ii) a way to construct the 2D kinematic maps of GCSs,
and (iii) the architectures of CNNs used in the present study
in Section 2.
We present the results of the predictions from CNN for disk
and elliptical galaxies, in particular, the prediction accuracies of CNNs
in the mass estimation in Section 3.
Based on these results,
we provide several implications of the present results
in the context of mass estimation of galaxies using GCSs of galaxies
in Section 4.
We summarize our  conclusions in Section 5.
In this paper, we focus exclusively on
the accuracy of prediction from CNNs in the mass estimation of galaxies
(including dark matter) from GCS kinematics . Accordingly, we do not
discuss other key issues related to the origin of GCSs in galaxies. Since we focus on the mass estimation of galaxies based on recent methods using GCS kinematics, we do not discuss other classic methods using the projected mass estimator,  \citep[e.g.,] []{ Bahcall1981} and the total numbers of bright GCs, \citep[e.g.,] []{Prole2019}.

\section{The model}

In this paper, we numerically examine the dynamical evolution of GCSs in isolated disk galaxies (``disk'' models)
and elliptical galaxies formed by major merging (``elliptical'' models).
We do not discuss the details of the physical characteristics of the disk and elliptical galaxies as we focus exclusively on the structures and kinematics of
GCSs. The details of the evolution of the interstellar medium (ISM) are highly unlikely to influence the physical properties of GCSs because GCs are mostly located outside the gaseous and stellar disks of disk galaxies. We investigate isolated disk galaxy models with different rotation and different ratios of stellar bulges to the stellar disk of their GCSs. For merger models, we consider merging of two spiral galaxies with various bulge-to-disk-ratios and baryonic mass fractions. In our previous simulations, we investigated the chemodynamical evolution of disk and elliptical galaxies using our original simulation codes (i.e. Bekki \& Shioya 1998; Bekki 2013). We use our original simulation code developed by \citep{Bekki2013}, though we do not investigate the evolution of gas, metals, and dust in the present paper.

\subsection{Structure and kinematics of a stellar disk}

The total masses of dark matter halo, gas disk, stellar disk, and
bulge of a disk galaxy are assumed to have masses of $M_{\rm h}$, $M_{\rm g}$, $M_{\rm s}$, 
and $M_{\rm b}$, respectively.
In this preliminary work, we only show the results 
of models with no gas
($f_{\rm g}=M_{\rm g}/M_{\rm s}=0$).
The ratio of stellar bulge to stellar disk is 
denoted
by a parameter $f_{\rm b}$ and is determined as $M_{\rm b}/M_{\rm s}$. The fundamental parameters in this investigation
are $M_{\rm h}$, $f_{\rm bary}$, and $f_{\rm b}$. We assume that the radial density profile of the dark matter halo is described as Navarro, Frenk \& White model suggested from Cold Dark Matter simulations in order to describe the initial density profile of dark matter halo in a disk galaxy \citep[]{Novarro1996}. It is explained as:

\begin{equation}
{\rho}(r)=\frac{\rho_{0}}{(r/r_{\rm s})(1+r/r_{\rm s})^2},
\end{equation}
where $\rho_{0}$ is the characteristic density of a dark halo, $r$ is the spherical radius and $r_{\rm s}$ is defined as the scale length of the halo. The ratio of $r_{\rm vir}$ and $r_{\rm s}$ ($c=r_{\rm vir}/r_{\rm s}$) is known as $c$-parameter, where $r_{\rm vir}$ is known as the virial radius of a dark matter halo. The value of $r_{\rm vir}$ for a given dark halo mass ($M_{\rm dm}$) is chosen by using the $c-M_{\rm h}$ relation for $z=0$, \citep[e.g.,][]{Neto2007}.

%%% TABLE 1
\begin{table}
\centering
\begin{minipage}{80mm}
\caption{Description of the basic parameter values
for the fiducial model for merging of two disk galaxies
}\label{Table 1}
\begin{tabular}{ll}
{Physical properties}
& {Parameter values}\\
Total halo mass (galaxy)
& $M_{\rm dm}=1.0 \times 10^{12} {\rm M}_{\odot}$  \\
DM structure (galaxy)
& NFW profile \\
galaxy virial radius (galaxy)
&  $R_{\rm vir}=245$ kpc  \\
$c$ parameter of galaxy halo
&  $c=10$  \\
Number of GCs in a GCS & $N_{\rm gc}=200$     \\
Rotational energy fraction of a GCS & $f_{\rm rot}=0.0$     \\
Stellar disk  mass & $M_{\rm s}=6.0 \times 10^{10} {\rm M}_{\odot}$     \\
Stellar disk size & $R_{\rm s}=17.5$ kpc \\
Gas disk size & $R_{\rm g}=17.5$ kpc \\
Disk scale length & $R_{0}=3.5$ kpc \\
Bulge mass &   $M_{\rm b}=10^{10} {\rm M}_{\odot}$  \\
Bulge size  & $R_{\rm b}=3.5$ kpc \\
Mass resolution & $5.0 \times 10^4 {\rm M}_{\odot}$ \\
Size resolution & 175 pc \\
Mass-ratio of two spirals in a merger & 1.0 \\
Initial distance of two spirals in a merger & 140 kpc \\
Circular velocity factor of a merger  & $f_{\rm v}=0.45$  \\
Star formation  & Not included \\
Chemical  evolution  &  Not included \\
Dust  evolution  &  Not included \\
\end{tabular}
\end{minipage}
\end{table}

\begin{figure*}
    \centering
    \includegraphics[width=18 cm]{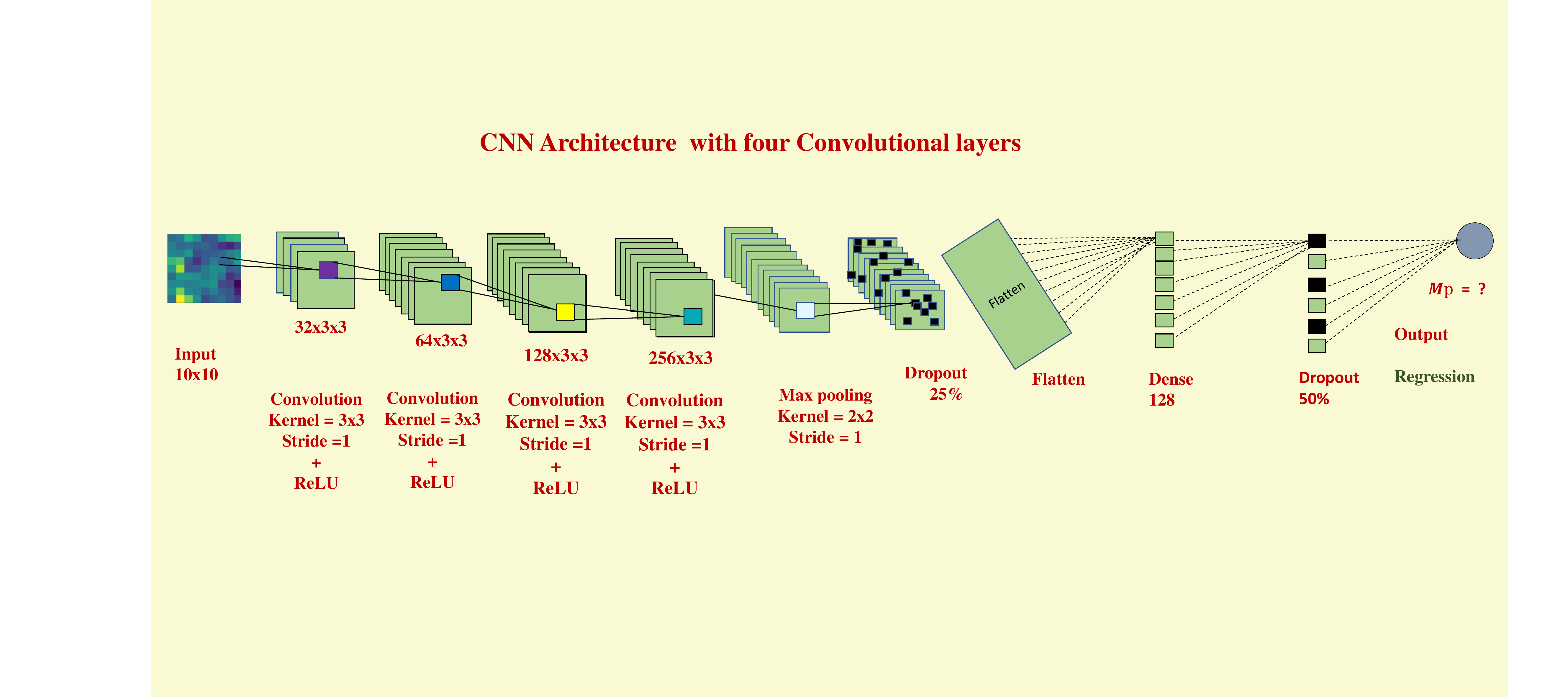}
    \caption{The adopted architecture of our original CNN with four convolutional layers followed by max-pooling layer, dropout and dense layers and a linear output layer. The details of the architecture are explained in the main text.}
    \label{CNN Architecture}
    
\end{figure*}

The Hernquist
density profile represents the bulge of a disk galaxy. This bulge of disk galaxy has a scale-length of $R_{\rm 0, b}$ and
size of $R_{\rm b}$. The bulge is supposed to have radial and isotropic velocity dispersion which are explained by Jeans equation
for a spherical system.
The value of bulge-to-disk ratio ($f_{\rm b}=M_{\rm b}/M_{\rm d}$) for a disk galaxy ranges from 0 (pure disk galaxy) to 4 (bulge-dominated).
The `Milky Way-type' models are those
with $f_{\rm b}=0.17$ and $R_{\rm b}=0.2R_{\rm s}$,
where $R_{\rm s}$ is the stellar disk size of a galaxy.
We use the mass-size scaling relation which is given by  $R_{\rm b} = C_{\rm b} M_{\rm b}^{0.5}$. From this relation, $R_{\rm b}$ for a given $M_{\rm b}$ can be determined. The value of $C_{\rm b}$ is determined so that $R_{\rm b}$ can be 3.5 kpc
for $M_{\rm b}=10^{10} {\rm M}_{\odot}$ (corresponding to the mass and size
of the Milky Way's bulge).

The vertical ($Z$) and radial ($R$) density profiles of the stellar disk are proportional to $\exp (-R/R_{0})$ having scale
length $R_{0} = 0.2R_{\rm s}$  and ${\rm sech}^2 (Z/Z_{0})$ with scale
length $Z_{0} = 0.04R_{\rm s}$ respectively.
The model we use for our study has $R_{\rm s}=17.5$ kpc. 
The initial radial and azimuthal velocity dispersion are assigned to the disc component using the epicyclic theory with Toomre's parameter, $Q$ = 1.5, along with the rotational velocity produced by the gravitational field of the disk. The vertical velocity dispersion at a given point is set to be half the radial velocity dispersion at that point.

The total number of particles in a fiducial model with $f_{\rm b}=0:167$
is $1016700$ and it depends on $f_{\rm b}$. The mass resolution
is $1.2 \times 10^4 {\rm M}_{\odot}$ in all models of the current study.
The gravitational softening length for every component is decided by the size of distribution and a total number of particles in each component ($R_{\rm s}$ and $r_{\rm vir}$). It is set to be 320 pc, which is much
finer than $1-2$ kpc spatial resolutions used for the image analysis in this study.

These spatial and mass resolutions are not particularly high, mainly because
we have to run numerous models for the limited amount of
computing time allocated
for this project. We believe that the adopted resolutions are good enough
to construct the CNNs (later described) and analyze the image data from
simulations.

\subsubsection{Globular cluster systems}

We briefly discuss the model as we use the same model for GCSs as used in \citep{Bekki2005}. The GCS in a disk galaxy is supposed to 
have a radial density profile of ${\rho}(r)$ $\propto$ $r^{-3.5}$.
Therefore, we assume that a GCS in a disk has radial density profile given by following equation:
\begin{equation}
 {\rho}(r)=\frac{\rho_{gc,0}}{({a_{\rm gc}}^2+r^2)^{1.75}},
\end{equation} 
where $r$ is the spherical radius, $\rho_{\rm gc, 0}$ is the central number density of the GCS, and $a_{\rm gc}$ is the scale
length of the GCS.

This $\rho_{\rm gc, 0}$ is defined according to the $a_{\rm gc}$.
In this study, we do not distinguish between metal-poor and metal-rich GCSs, though such two different GC populations are considered in \citep{Bekki2005}.
Therefore, we adopt a single value of $a_{\rm gc}=0.3R_{\rm d}$ ($\sim 5$ kpc for a Milky Way-type disk galaxy) for a GCS. This adopted value is consistent with the observed GC distribution of the galaxy. The cut off radius ($R_{\rm c}$) 
is set to be $3R_{\rm d}$ for GCSs ($\sim 50$ kpc for a Milky Way-type disk galaxy). This is a radius beyond which no GC particles are initially allocated.

The GCS in a disk is supported by both rotation velocity and velocity dispersion. This velocity dispersion is assumed to be isotropic.
We introduce a parameter $f_{\rm rot}$ that describes the ratio of
total rotational energy 
($T_{\rm rot}$) of a GCS to its total kinetic energy ($T_{\rm kin}$):

\begin{equation}
f_{\rm rot}= \frac{ T_{\rm rot} }{ T_{\rm kin} } .
\end{equation}

We assume that each GC is rotating around the spin axis of its host
galaxy (the $z$-axis) and the above $f_{\rm rot}$ determines its rotational velocity. In order to estimate the rotational velocity
of each GC, we take the following steps.

We first evaluate the 3D velocities of a GC particle from the gravitational potential
at the position of dynamical equilibrium. This is a position where velocity dispersion at each GC should be
the same as that for dark matter. We then add the rotational velocity to
the GC and reduce the 3D velocities so that $f_{\rm rot}$ can be the adopted value.
The total number of GCs in a disk galaxy is 200.

\subsubsection{Orbital configurations of galaxy merging}

The orbits of the two disks are initially set to be in the $x$-$y$ plane in all the merger simulations with different mass-ratios of two disks
($m_2$). The distance between the centers of mass of the two disks is $[4-8]R_{\rm d}$ for all models. The relative velocity of the two galaxies is $v_c*f_v$, where $v_c$ is the circular velocity at the initial distance and $f_v$ is the circular velocity factor ($f_v$) ranging from $0.05$ (highly radial) to $1.0$ (circular orbit). The GCSs of merger remnants in the models with larger $f_v$ can show higher degrees of global rotation (i.e., More angular momentum in the GCSs), because the initial two disk galaxies can have larger orbital angular momentum. On the other hand, the models with small $f_v$ can have more radial orbital of GCs and lower intrinsic angular momentum in GCSs. Therefore, the diverse kinematics in GCSs can be seen in these models with different $f_v$.

$\theta_i$ denotes the spin of each galaxy in a merger, where the subscript $i$ identifies each galaxy.  Here, $\theta_i$ is the angle between the vector of the angular momentum of the disk and the $z$-axis. We change these angles for the two disks randomly. The azimuthal angle $\phi$ is 
measured from the $x$-axis to the projection of the angular momentum vector of the disk onto the $x$-$y$ plane, and this is set to be 0 for the two disks because this is not an important parameter in the present study.
The time when 
the progenitor disks merge entirely and reach the dynamical equilibrium is
less than 20.0 in our units for most of the models,
though it is longer for smaller $m_2$.

\subsubsection{Method to derive 2D kinematics maps of GCS}

In order to produce 2D maps of GCSs in a simulated galaxy,  we first divide the 1.2 Rd x 1.2 Rd square
region of the galaxy  (with the center being consistent with the mass center) into 10 x 10 meshes (pixels) and thereby count the number of GCs in each mesh. Then we estimate the mean line of sight velocity and velocity dispersion for each mesh, though the number of GCs in each mesh is not large owing to the adopted total number of GCs (200 for each galaxy). The Gaussian smoothing method adopted in our previous works \citep[]{Bekki2019, Diaz2019} is
applied to these 2D kinematics maps of GCSs.
 Fig. 2 shows an example of the estimated 2D kinematics maps for an isolated disk galaxy model. For each galaxy model, 2D kinematics maps are generated for each of 100 viewing angles. Therefore, if we run 100 merger models with different orbital parameters and mass-ratios of mergers, then 10,000 images can be generated. 80\% of these are randomly selected for training and 20\% of them
are used for testing the proposed CNN model.

Additionally, we also developed new dataset from galaxy models for testing of the proposed CNN model which is different to the above mentioned data. This dataset is referred to as ``completely unknown’’ dataset in this study, because the model parameters of galaxy merging (e.g., the mass-ratios of galaxy mergers) are different from those used for the above-mentioned dataset. The structures and kinematics of GCSs in the completely unknown dataset can be different from those of the above dataset so that we can make a more robust and thorough testing for our proposed CNN model. The results for this completely unknown dataset is described in Section 3.5.

In this study, we propose a simple CNN model trained by simulation dataset and tested by simulated as well as real observational data
from Alabi et al. (2016,A16) to discuss whether or not our proposed CNN model can accurately predict the total masses of 
galaxies from their GCSs. In order to make a fair and consistent comparison between the predicted total
mass of a galaxy from our model and the observationally estimated one by A16, we tried to generate 2D kinematics maps from observational data in the same way. We use the latest observational result obtained by Dolfi et al. (2020, D20) for the nearby early-type galaxy NGC 3115, which could be a remnant of major merging. We divide the observed GCS of NGC 3115 into
10 x 10 meshes and estimate the mean velocity and velocity dispersion for each mesh, as we do for the simulated galaxies. The details of this comparison are presented in Section 4.1.

\subsection{CNN Architecture}

%%%%% TABLE1 
\begin{table*}
\centering
\begin{minipage}{180mm}
\caption{A brief summary for the adopted architectures of CNNs and the results for dataset from (i) isolated disk models, (ii) elliptical galaxies models, and (i)+(ii).}\label{Table 2} 
\begin{tabular}{llllllll}
Name
& Number of images 
& Model image
& CNN architecture
& Training epochs
& Accuracy (One-Channel)
& Accuracy (Two-Channel)\\
DL1 & 100,000 & disk galaxy & 2 convo  & 1000 & 0.937 & 0.965 \\
DL2 & 80,000 & merger & 2 convo  & 1000 & 0.972 & 0.980 \\
DL3 & 200,000 & disk + merger & 2 convo  & 1000 & 0.946 & 0.959 \\
DL4 & 200,000 & disk + merger & 4 convo & 1000 & 0.976 & 0.978 \\
DL5(noise) & 200,000 & disk + merger & 4 convo & 1000 & 0.809 & 0.812 \\

\end{tabular}
\end{minipage}
\end{table*}

\begin{figure}
    \includegraphics[width=9cm]{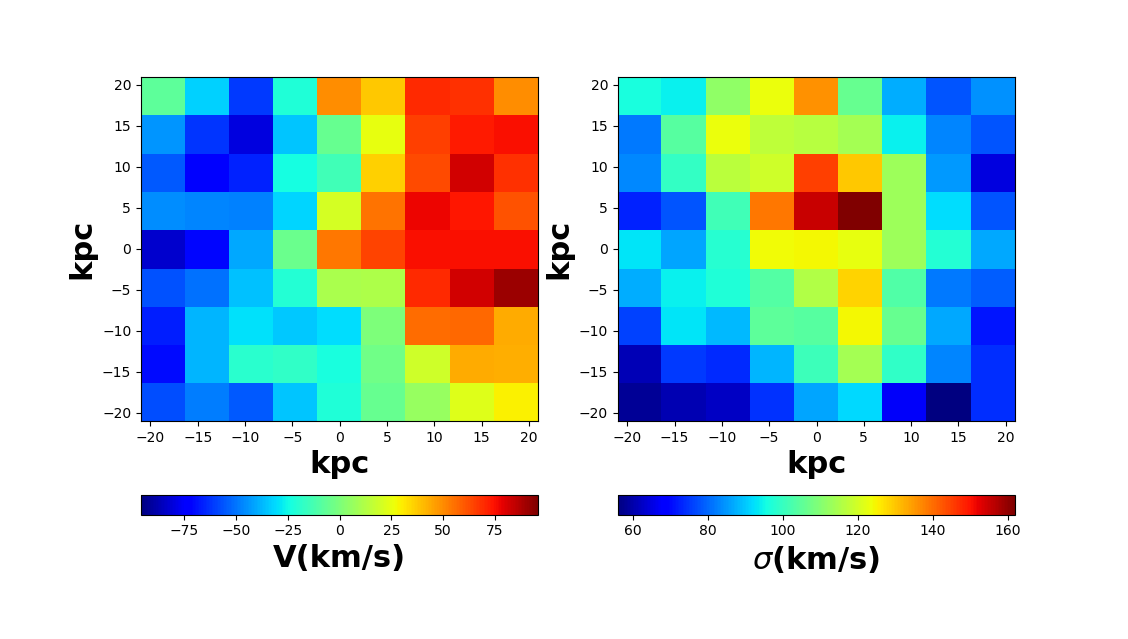}
    \caption{2D maps of line-of-velocity ($V$, left) and  velocity dispersion ($\sigma$, right) in the isolated disk model with global rotation in its GCS. The maps are derived within $R<1.2$ $R_d$, where $R_d$ is the initial size of the stellar disk in this model.}
    \label{2d heatmap}
\end{figure}

\begin{figure}
    \includegraphics[width=8cm]{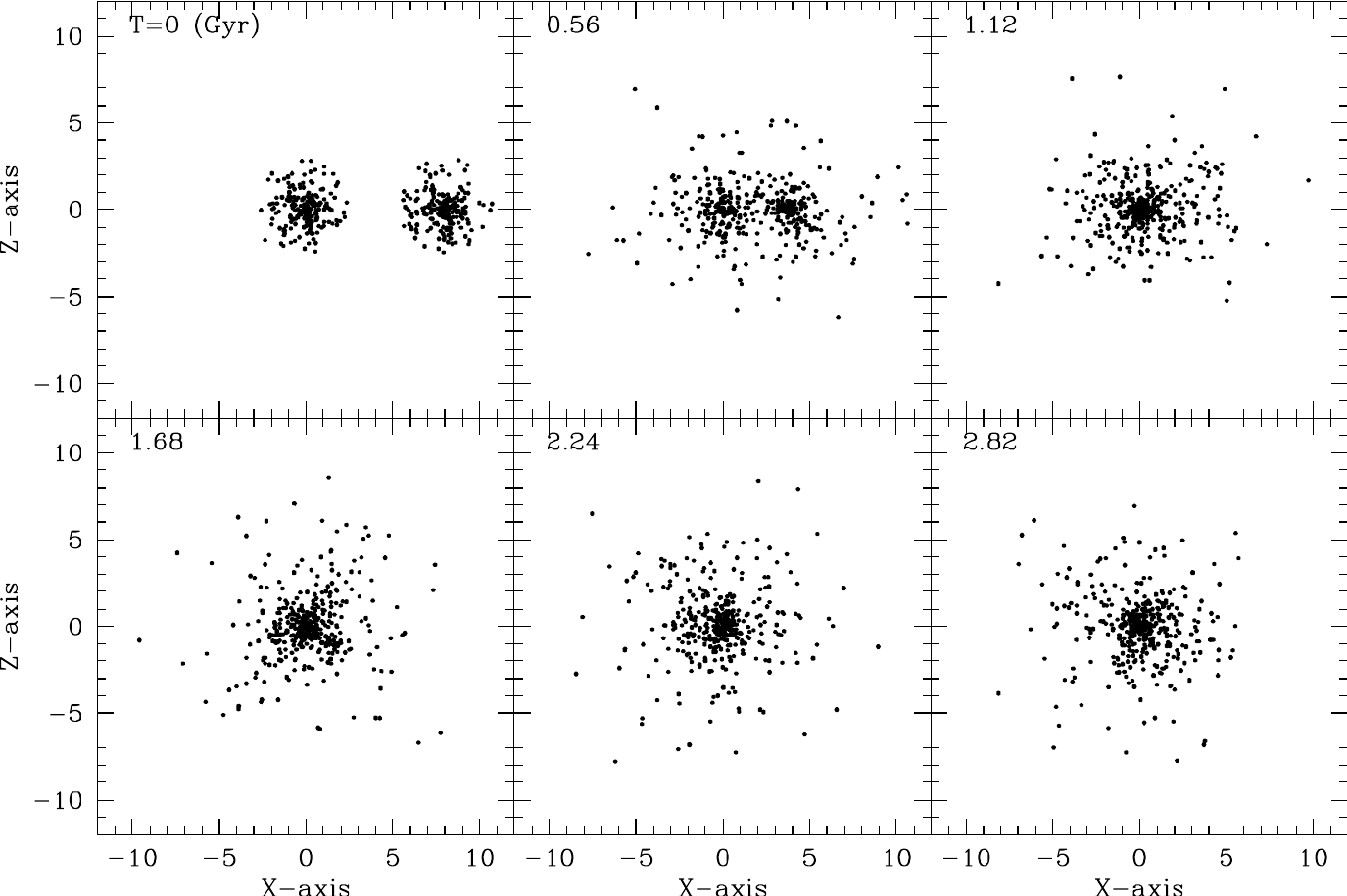}
    \caption{Time evolution of the projected distribution of GCs onto the $x-z$ projection (the orbital plane of the merger) in the fiducial merger model with $f_v=0.45$ and $N_{gc}=200$. Each dot is a  GC, and the scale is given in simulation units (17.5 kpc).}
    \label{Evolution of galaxy clusters}
\end{figure}

\begin{figure*}
    \includegraphics[width=19cm]{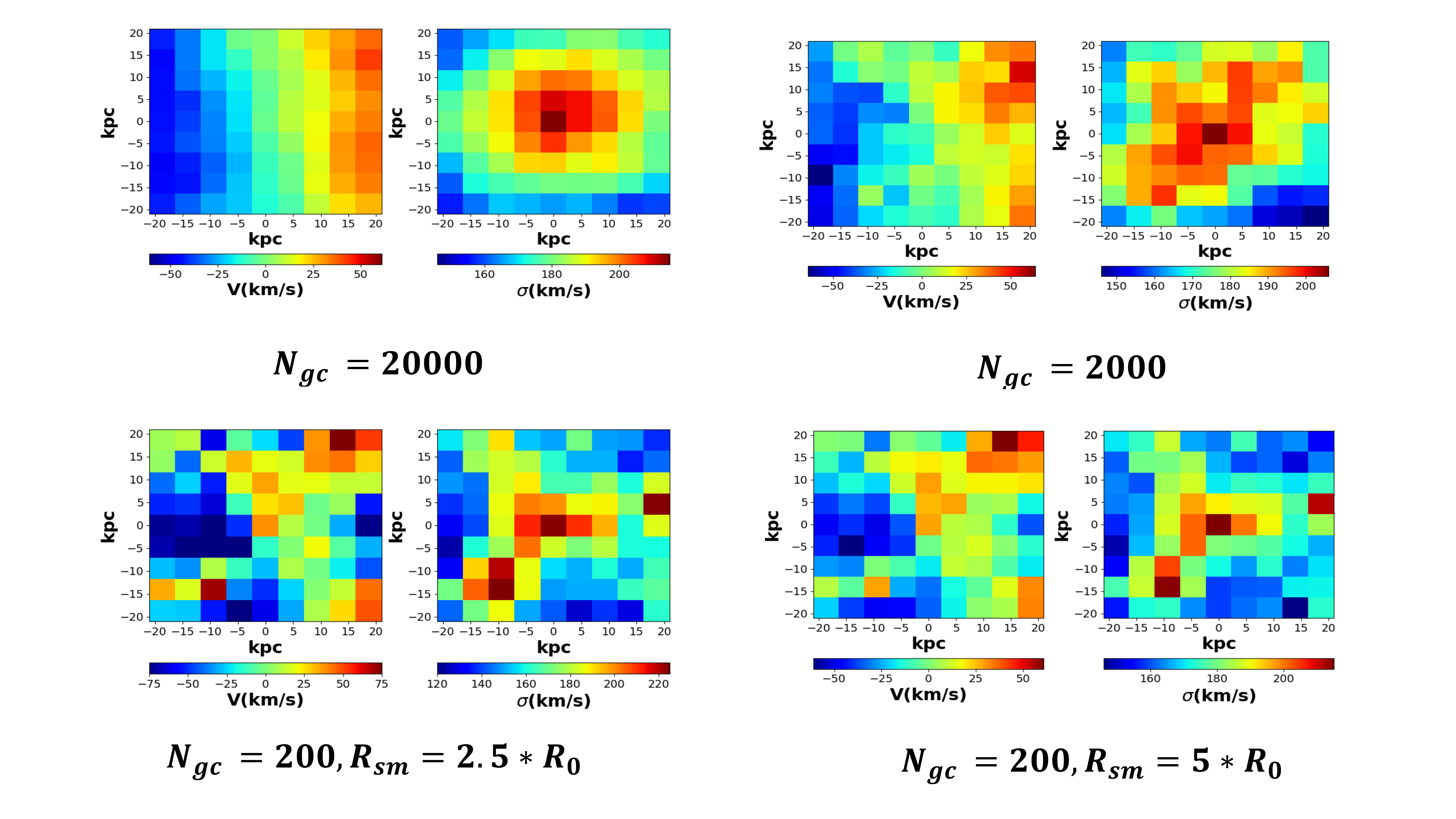}
    \caption{2D maps of $V$ and $\sigma$ for merger models with $N_{gc}=20000$ (upper left), $N_{gc}=2000$ (upper right), $N_{gc}=200$ and $R_{sm}=2.5 \times R_0$ (lower left) and $N_{gc}=200$ and $R_{sm}=5 \times R_0$, Here $R_0$ is the scale length of the initial stellar disk.}
    \label{combined image}
\end{figure*}

In the literature, CNNs have become the dominant deep learning approach for visual object recognition. It is used for image classification, object detection and segmentation. CNNs are not restricted to visual perception and are also successful at other tasks, such as voice recognition and natural language processing.

In the present study, we will use CNNs with the architectures similar to those used in our previous study for galaxies under ram pressure stripping and galaxy interaction for the regression problems of the present study \citep[]{Bekki2019,Diaz2019,Cavanagh2020}. We also use a CNN that is newly developed for the purpose of more accurate predictions of total masses of galaxies from GCSs.

The CNN network that we used for our model is summarized in Figure \ref{CNN Architecture}. In the following subsections, we describe the overall architecture of our CNN model.  

\subsubsection{Convolutional layers}

The architecture initially consists of two convolutional layers. Neurons in the first convolutional layer are only connected to pixels in their receptive fields. Similarly, each neuron in the second convolutional layer is connected only to the neurons in the receptive field of the first convolutional layer. This architecture enables the network to focus on low-level features in the first hidden layer, then combine them into higher-level features in the next hidden layer \citep[]{Geron2017}. 

The first convolutional layer filters the $(10\times 10\times 1)$ input image with 32 filters of size $(3\times 3\times 1)$ having stride of length 1 (both horizontally and vertically) and ``valid'' padding. Here, the stride is a distance between two receptive fields. In convolutional layer, choosing an appropriate number of filters is significant for feature detection. We can start any neural network with filter range [ 32, 64, 128 ] in the input layers and increasing their size to [ 256, 512, 1024 ] in deeper layers \citep[]{Geron2017}.

The convolutional layers require another parameter which is kernel size. There are various kernel sizes available $(1,1),(2,2),(3,3),(5,5)$ and $(7,7)$. Since our input size for the images is $(10\times10)$ pixels, we prefer to use the kernel size of $(3\times 3)$. If we have a larger image, then we can use larger kernels. Also encouraged by the architecture of VGGNet, we decided to use the kernel size of $(3\times 3)$ \citep[]{Simonyan2014}. Similarly, the second convolutional layer in our network has 64 filters of the same kernel size $(3\times 3)$ with stride $(1\times1)$. The number of parameters in the first and second convolutional layer are 320 and 18,496, respectively.

Moreover, we used our convolutional layers with ReLu activation function, because ReLu trains network several times faster as compared to the tanh activation function (Krizhevsky,  Sutskever \& Hinton 2017). The study showed that a four-layer CNN with ReLUs attained 25\% training error rate on CIFAR-10 (i.e. it took 6 times less time to train than a comparable network with a tanh activation function ) \citep[]{Krizhevsky2017}. Therefore, we decided to select ReLU as an activation function with each convolutional layer to save computing time.

\subsubsection{Pooling layers}
Max-pooling layer in CNN architecture offers stronger translation invariance than average pooling, and it is comparatively computationally efficient. Therefore, max-pooling layers are now more commonly used as compared to average pooling layers. Hence, we decided to use only one max-pooling layer with $(2\times2)$ and stride of 1 (both horizontally and vertically) in order to get the best features and reduce computational complexity. It is used to reduce the input shape further, and the network does not learn anything from this layer. Therefore, the number of parameters in max-pooling remains 0.

\subsubsection{Dropout}
Dropout has proven to be highly successful and is reported to give 1-2\% accuracy improvement in results \citep[]{Hinton2012}. The dropout technique consists of producing zero output for each hidden neuron with some probability. We use dropout with a fully connected layer with probability 0.50. The purpose of this layer is to help our network generalize and not to over-fit. Neurons from the current hidden layer with probability $p$ will randomly disconnect from neurons in the next hidden layer so that the network has to rely on the existing connections. The hyper-parameter $p$ is called the \textit{dropout rate}, and it is typically set to 50\%. Since dropout just removes the nodes that are below the mentioned weights, the number of parameters in the dropout layer remains zero. At the testing phase, we simply scale the weights by the chosen dropout rate and use the network for making predictions \citep[]{Nitish2014}.

In our architecture, we used two \textit{dropout} layers; one is on top of max-pooling layer while the other is on top of the dense layer. We set our max-pooling layer to have a \textit{dropout rate} of 25\%, and a dense layer with 128 neurons having a drop out rate of 50\%.

\subsubsection{Output layer}
The last layer of our proposed CNN architecture is the output layer containing 129 parameters (128 connections coming in from previous dense layer + 1 bias neuron). The formula to calculate parameters at this layer is given by:

\begin{equation}\label{errors}
 N = ((c\times p) +1 \times c)
\end{equation}
where N is the number of parameters in current layer, c is the number of neurons in current layer, p is the number of neurons in previous layer and 1 is used for bias neuron.

We do not use any dropout layer on the output layer. In addition to this, a ``linear activation function'' is added at the last layer to the network. The output of this layer is the prediction value itself. In our case, its output is the mass of the dark matter $M_p$ which is normalized to 0 to 10. We have converted it into physical units for the output.

\subsubsection{CNN with four convolutional layers}

One of the most critical choices in network design was the number of convolutional layers. It was noticed that successive 2D convolutional layers have the effect of transforming input to extract increasingly high-level representation, at the expense of spatial resolution \citep[]{Hariharan2017}. In our research, we tested our CNN with a different number of convolution layers,  to investigate the trade-off between extracting higher-level features and preserving spatial resolution. Initially, we started with two layers and tested our model on one-channel and two-channel data. Then, motivated by AlexNet's architecture \citep[]{Alex2012}, we put four convolution layers directly on top of each other. Table \ref{Table 2} and Table \ref{MAE & RMSE} show the results with both architectures. We found that the accuracy of our model increased by adding more convolution layers. Therefore, our final model contains four convolutional layers on top of each other as shown in Fig. \ref{CNN Architecture}.

 The third convolutional layer has 128 filters with kernel size 3x3 followed by next convolutional layer having 256 filters with the same kernel size 3x3. The total number of parameters in the architecture with two convolutional layers is 92,930. However, the number of parameters significantly increases to 420,994 in case of four convolutional layers architecture.

\subsubsection{Performance comparison measures}

We have calculated errors to compare the performance of our model on one-channel and two-channel data using error density plots, given by the following equation:

\begin{equation}\label{errors}
 E_i=\frac{(M_p-M_c)}{M_c}
\end{equation}
where, ${M_p}$ and ${M_c}$ represent the predicted and actual mass of dark matter, respectively.

Moreover, it is also necessary for our research to look at cumulative distribution plots in order to compare the performance of one-channel and two-channel data. We use the absolute values of errors to plot cumulative distribution plots as given in the following equation:

\begin{equation}\label{abs errors}
 |E_i|=\frac{|M_p-M_c|}{|M_c|}
\end{equation}

Additionally, we use the well-established Root Mean Square Errors (RMSEs) and Mean Absolute Errors (MAES) as metrics for evaluating model performance. We have also used accuracy of the model in order to compare performance of proposed model for images from different galaxies. $M_c$ and $M_p$ are the normalized values ranging from 0  to 10. We converted them into Msun for the output, but, the real DM masses are used for the accuracy estimation. Moreover, the number of bins used for accuracy estimation is 20. In the following section, we will describe our significant results of estimation of dark matter mass using the proposed CNN architecture in detail.

%%%%% TABLE1 
\begin{table*}
\centering
\begin{minipage}{180mm}
\caption{Summary of results for CNN model used for disk, merger and disk + merger galaxies.}\label{MAE & RMSE}
\begin{tabular}{lllllllll}
Name
& Images 
& Model Image
& MAE (One-channel)
& MAE (Two-channel)
& RMSE (One-channel)
& RMSE (Two-channel)\\ 
DL1 & 100,000 & disk galaxy & 0.571 & 0.404  & 0.833 & 0.707 \\
DL2 & 80,000 & merger & 0.265  & 0.188  & 0.357 & 0.239  \\
DL3 & 200,000 & disk + merger & 0.501 & 0.367 & 0.733 & 0.595  \\
DL4 & 200,000 & disk + merger & 0.288 & 0.275  & 0.539 & 0.51 \\
DL5(with noise) & 200,000 & disk+merger & 2.722 & 2.835 & 3.234 & 3.408 \\

\end{tabular}
\end{minipage}
\end{table*}

\section{Results}
We first briefly show the 2D kinematics maps of GCs in the fiducial merger model with different $N_{gc}$, because we need to demonstrate how the present results can depend on $N_{gc}$ in Section \ref{sec:3.1}. Then we describe the results of the CNN-based mass from GCSs in 
disk galaxies using 100,000 images dataset (e.g, 2D maps of $\sigma$ and $V$) in Section 3.2.
Table \ref{Table 2} describes the number of images and the model parameters
for CNNs.
For these disk galaxy models,
we divide the images  into 80,000 for training CNNs and 20,000 for testing.
We also describe the results for the elliptical galaxy models (i.e.,
galaxy mergers) with
80,000 images in Section 3.3. 
We then show the results for the combined dataset from disk and elliptical
galaxy models in Section 3.4. For these results, the testing data are different from
the training one, but both are generated from the same galaxy models. In Section 3.5,
we test the CNN trained by these images for the totally unknown dataset that
is not used in training CNNs. Thus we can check the prediction accuracy
of our CNNs in a more stringent manner. 
In Section 3.6, we demonstrate how the obtained results depend on
the CNN architectures by comparing the prediction accuracies between
different CNNs.

\subsection{2D Kinematics maps for GCSs}\label{sec:3.1}
We generated 100 images of the GCS in a galaxy viewed from 100 different angles. Therefore, one galaxy has 100 different images. For example, DL1 model sequence has 100,000 images for training, which means 100,000/100 = 1000 different GCSs (galaxies).
Since the number of GCs are quite different in different elliptical galaxies, 
we investigate the details of 2D maps for the present study to investigate 
how the 2D maps of $V$ and $\sigma$ depend on the number of GCs ($N_{gc}$). 
Fig. \ref{combined image} describes the 2D maps 
for the fiducial merger model with different $N_{gc}= 
200, 2000$ and $20000$ and different Gaussian 
smoothing length of $R_{sm}= 2.5R_0$ and $5.0R_0$ for $N_{gc}=200$.
This demonstrates that (i) global rotation of this GCS can be clearly seen
in all four models, i.e., even in the model with a low GC number
and (ii) the radial gradient of $\sigma$, which is a key in the mass estimation
of galaxies, can also be clearly observed.
These details of GCS kinematics,
such as radial gradients of $\sigma$ and $V$ profiles,  are the 
characteristic features that our CNNs can capture to  estimate
the total masses of galaxies. 

However, the random noise due to a small number of GCs per pixel can be more
remarkable for the models with low GC numbers. It is a key question in this
study whether the trained CNN can predict accurately the total masses of galaxies
with low GC numbers (i.e., with nosier 2D maps).
As shown in the subsequent subsections, even our CNNs trained by
images from GCSs with $N_{\rm gc}=200$ can accurately predict the total masses
of galaxies from GCS kinematics. This implies that noisy images do not greatly
prevent CNNs from learning the characteristic features of GCS kinematics that
depend on galaxy masses. Other problems related to observational noise 
in spectroscopic determination of $V$ and $\sigma$ will be discussed in Section 4.

\begin{figure*}
    \centering
    \includegraphics[width= 18 cm]{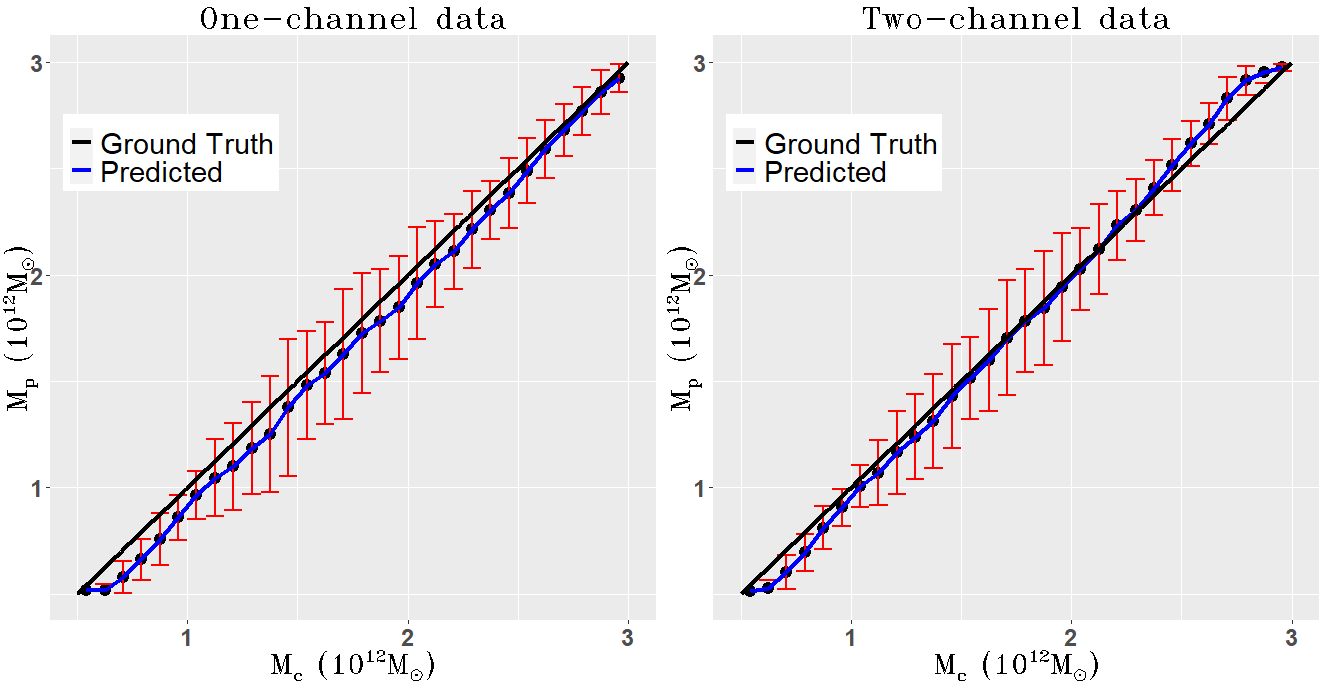} 
    \caption{Predicted vs correct dark matter mass within disk galaxies for one-channel (left) and two-channel data (right). Here one-channel (two-channel) means that only $\sigma$ data (both $\sigma$ and $V$ data) is used in the CNN-based prediction. $M_c$ on the x-axis represents the correct value of dark matter mass, and $M_p$ on the y-axis represents predicted dark matter mass. Error bars in the plot represent overall mean and standard deviation of the dark matter masses within disk galaxies. The Black line indicates $M_p = M_c$, and  the blue line represents the predicted mass values for dark matter within disk galaxies. \textbf{The number of bins used is 20.}}
    \label{Disk galaxies}
\end{figure*}

\subsection{CNN prediction of disk galaxies}
\label{sec:CNN}

In the first set of experiments, 
we have selected 100,000 images from disk galaxies, 80\% of images were randomly 
selected for training while rest of 20\% were dedicated for 
testing the performance of our network. 
We have trained our CNN for 1000 epochs for disk galaxies models and for
others in this study, because 1000 epochs are enough for the prediction
accuracy to become very high (more than 95\%),
From Table \ref{Table 2}, the model's accuracy for one-channel and 
two-channel data for disk galaxies are 0.937 and 0.965, respectively. 
This indicates that the CNN trained
by two-channel data has a higher accuracy as compared to one-channel data. This is mainly because the CNN trained by two-channel
data can properly consider global rotation (i.e., rotational energy of
self-gravitating systems) in the mass-estimation. 

Furthermore, Fig. \ref{Disk galaxies} 
presents a detailed  comparison between the predicted galaxy mass and the correct
one   
(``ground truth'') for disk galaxies.
In this Fig. \ref{Disk galaxies}, 
$M_c$ on the x-axis represents correct 
dark matter masses and the y-axis 
indicates predicted dark matter mass 
in disk galaxy images. 
The error bar in each mass bin represents the overall 
mean and standard deviation for the predicted values of 
total galactic masses. 
The black line indicates where $M_c = M_p$ and the  blue one
connect the average of the predicted mass in each bin  in Fig. \ref{Disk galaxies} . 
This clearly demonstrates that the CNN can accurately predict the 
total mass of galaxies for a wide range of galaxy masses, though the standard
deviation is not very small.

It is observed from Fig. \ref{Disk galaxies} that
 the two-channel (velocity dispersion $\sigma$, and line-of-sight velocity $V$ of GCs) 
predictions are better as compared to one-channel (velocity dispersion $\sigma$) 
predictions as the error bars 
are small in case of two-channel data. 
For an idealistic behavior, the error-bar plot 
should have points symmetrically distributed 
around the 45-degree straight line. 
The two plots in Fig. \ref{Disk galaxies} show absolute symmetry. Hence, the accuracy of 93.7\% and 96.5\% are proved through these plots.

Moreover, Table \ref{MAE & RMSE} gives information about MAEs and RMSEs for one-channel and two-channel data for different galaxy models. Table \ref{MAE & RMSE} shows RMSEs for one-channel and two-channel data for disk galaxies are 0.833 and 0.707, respectively, and MAEs for one-channel and two-channel data are 0.571 and 0.404, respectively. This means model with two-channel data for disk galaxies performs better than the model with one-channel data as the values of RMSEs and 
MAEs for the model with two-channel data are smaller than the model with one-channel data.
These results again confirm that global rotation of GCSs should be considered for
more accurate mass estimation of galaxies.

\begin{figure}
    \centering
    \includegraphics[width=8 cm]{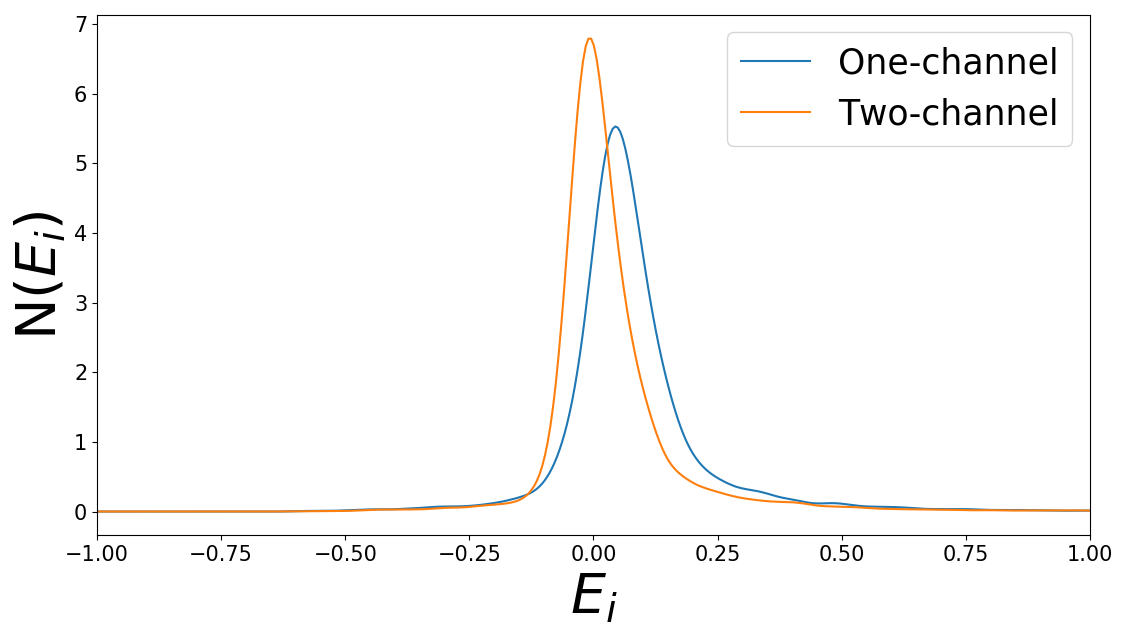}
    \caption{Distribution of errors $E_i$ in the mass estimation of dark matter halos within disk galaxies: one-channel (blue) and two-channel (orange). The ways to derive $E_i$ and $N(E_i)$ are described in the main text.}
    \label{Error distribution for disk galaxy}
\end{figure}

Furthermore, Fig. \ref{Error distribution for disk galaxy} presents the error distribution plot for one-channel and two-channel data for disk galaxy images. It is a line plot representing histograms. Since, the intervals on the x-axis are very fine, therefore, histograms are plotted as a line otherwise the plot will look cluttered
It can be observed that $E_i$ on the x-axis represents the 
errors calculated from Equ.~(\ref{errors}) using the predicted values of total galaxy mass and the ground truth (i.e., correct mass).
The y-axis represents the number of errors per each bin. 
Since the line of two-channel data (orange curve) 
near the zero is higher as compared to one-channel data 
(blue curve) in Fig. \ref{Error distribution for disk galaxy},
we can say that the model with two-channel data gives better 
predictions in comparison to the model with one-channel data in case of disk galaxies.  

\begin{figure}
    \centering
    \includegraphics[width=8 cm]{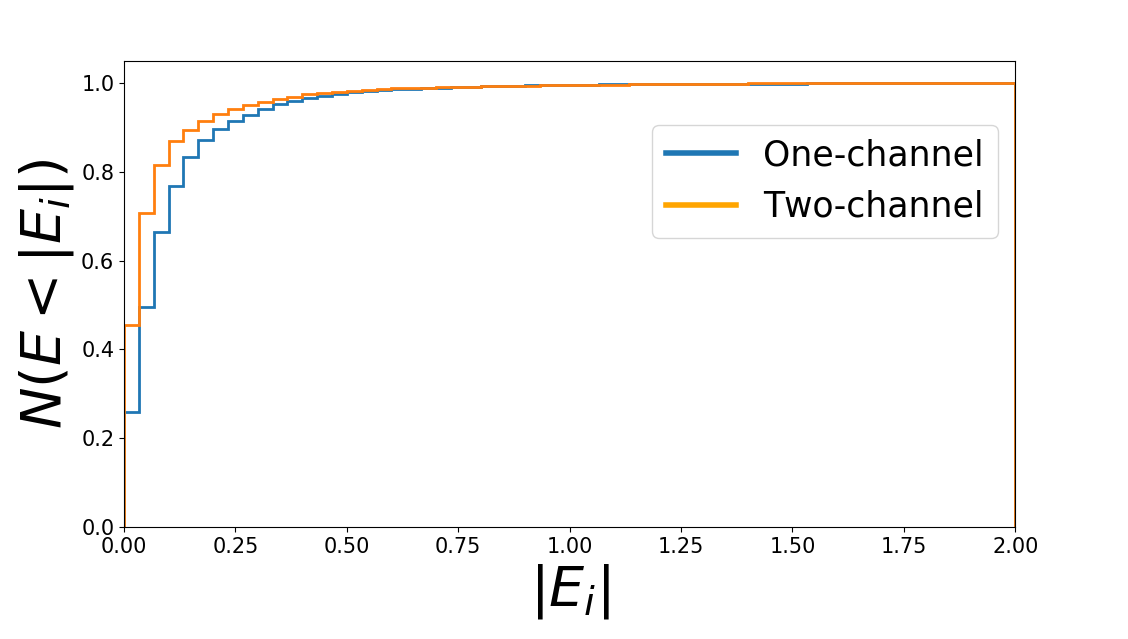}
    \caption{Cumulative error distribution graph for dark matter mass of disk galaxies in terms of physical units for one-channel (blue) and two-channel data (orange). Here, $|E_i|$ on the x-axis represents the absolute value of errors. The y-axis represents the normalized number of images per each $|E_i|$ bin.}
    \label{Cumulative error distribution graph for dark matter mass of disk galaxy}
\end{figure}

The performance of models with one-channel 
and two-channel data can be quantified reasonably well
by using the cumulative error distribution 
graph shown in Fig. \ref{Cumulative error
distribution graph for dark matter mass of disk galaxy}. 
In this figure, 
$|E_i|$ on the x-axis represents
the absolute values of errors calculated from Equ.~(\ref{abs errors}) using the predicted values of galaxy mass and the ground truth. 
The y-axis shows $N(E<|E_i|)$, i.e., the total number of images with the predicted $E$
less than a certain value ($E_i$).
 From Fig. \ref{Cumulative error distribution graph for dark matter
 mass of disk galaxy}, the model with two-channel data (orange curve) shows that around 
45\% of predicted dark matter mass has less than 5\% of errors. On the other hand, the model with one-channel data (blue curve) in the same Fig. \ref{Cumulative error distribution graph for dark matter mass of disk galaxy} shows more than 30\% of predicted dark matter mass has less than 5\% of errors. 
Hence, it is clear that our CNN 
model gives better predictions,
if the CNN is trained using two-channel data rather than one-channel data
for disk galaxies.

\subsection{CNN prediction for elliptical galaxies}
Our CNN is also trained by using  64,000 images from elliptical galaxies
formed from major merging with different orbital parameters and disk inclination
angles with respect to their orbital planes.
The predictions from our CNN for the  16,000 (20\% of data) testing images
are analyzed and  summarized in Figs. 8 -12. 
Table \ref{Table 2} shows that the model accuracy for elliptical galaxies using one-channel and two-channel data are 0.972 and 0.980, respectively, 
which is much better than our results for disk galaxies. 
The reason for this is as follows. The GCSs in disk galaxies are rotating along
the $z$-axis (i.e., the disk plans) so that global rotation can be properly captured
by our CNN if the galaxies are viewed edge-on.
Therefore, the two-channel predictions by our CNN work very well for such viewing 
angles.
However, if the galaxies are viewed from face-on, then the prediction accuracy
becomes worse, because the rotation factor in the mass estimation cannot be properly
considered by the CNN.
For the merger case, such viewing effects are less important so that
the mass estimation by our CNN can be more accurate.
The model with two-channel data performs better than the model with one-channel data for elliptical galaxies as well, as accuracy is higher in case of two-channel data.

\begin{figure}
    \centering
    \includegraphics[width=8 cm]{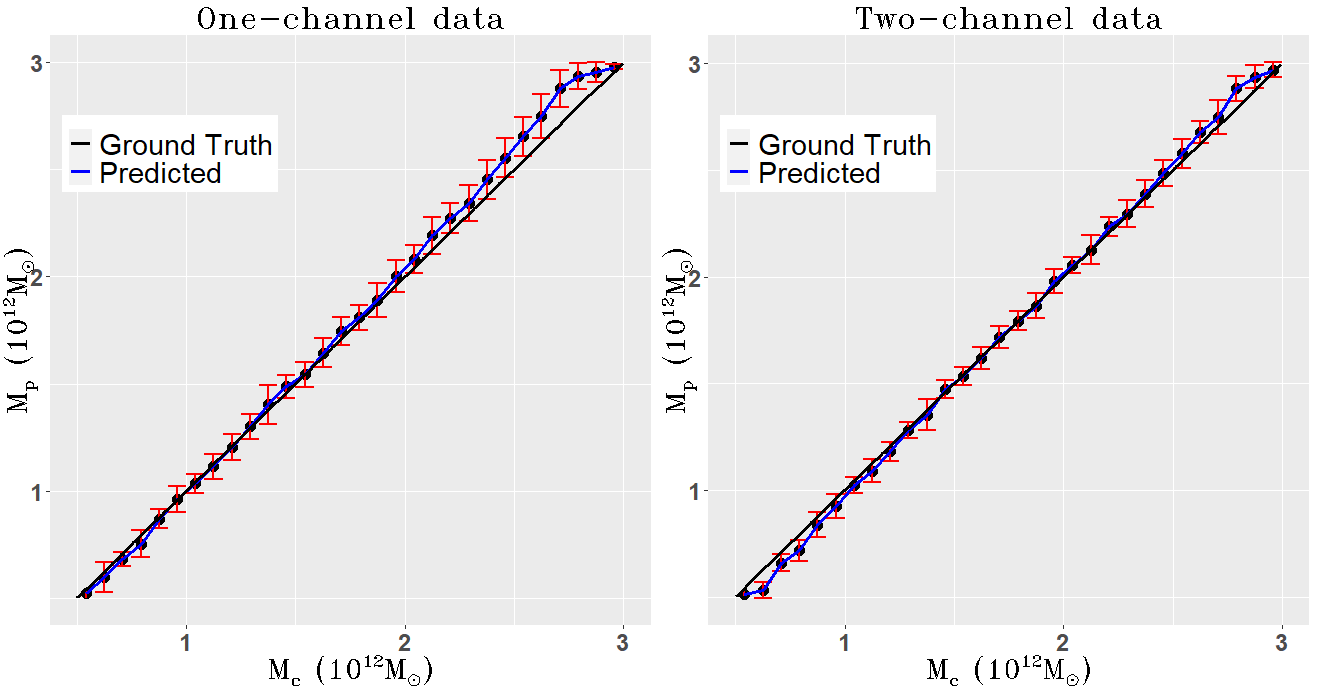}
    \caption{Predicted vs correct dark matter mass within elliptical galaxies for one-channel (left) and two-channel data (right).}
    \label{Merger galaxy dark matter mass in terms of Physical units}
\end{figure}

 Fig. \ref{Merger galaxy dark matter mass in terms of Physical units}
clearly demonstrates that our CNN's predictions are quite accurate 
for most images from elliptical galaxies with different total masses.
This confirms that 
 our CNN model gives better predictions in case of mergers 
as compared to isolated disk galaxies. 
The error-bar plot shows idealistic behavior as all the points 
symmetrically distributed around the 45-degree straight line. Hence, the accuracy of 97.2\% and 98.0\% for one-channel and two-channel respectively can be clearly observed in Fig. \ref{Merger galaxy dark matter mass in terms of Physical units}. 
The two-channel predictions in this case again look better as 
compared to one-channel predictions, which is totally consistent with the results from
other models
in this study
This again strongly suggests that rotational properties of GCSs should be properly
considered in the mass estimation of elliptical galaxies formed by major merging.

Table \ref{MAE & RMSE} 
shows RMSEs for one-channel and two-channel data for elliptical galaxies 
are 0.357 and 0.239 respectively, and MAEs 
for one-channel and two-channel data are 0.265 and 0.188 
respectively, for elliptical galaxies.  
Again in both cases, the model with two-channel 
data has smaller error values as compared to one-channel 
data (i.e., more accurate predictions in two-channel).
Moreover, these error values are almost half of the error values of disk galaxies,
which confirms that our CNN  performs better in case of elliptical galaxies as compared to disk galaxies.
Fig. \ref{Cumulative Error mergers} shows that (i) 
more than half of predicted total galaxy
mass has less than 5\% of errors and (ii)
the  two-channel CNN model gives better predictions in the elliptical galaxies too.

\begin{figure}
    \centering
    \includegraphics[width=8 cm]{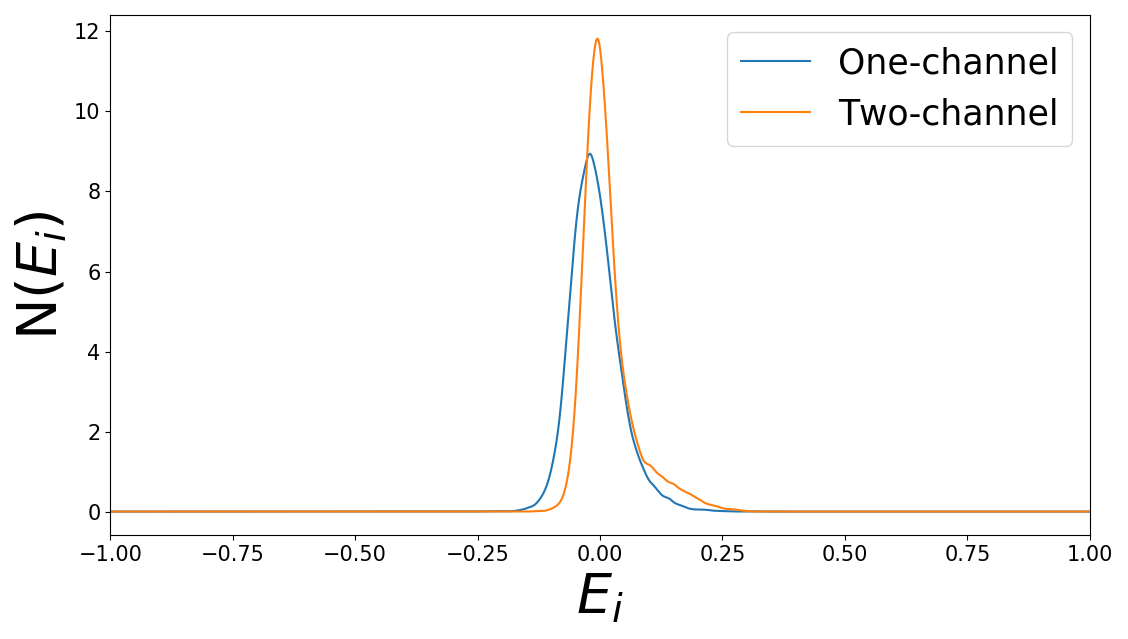}
    \caption{Distribution of errors $E_i$ in the mass estimation of dark matter halos within elliptical galaxies: one-channel (blue) and two-channel (orange).}
    \label{errors merger}
\end{figure}

\begin{figure}
    \centering
    \includegraphics[width=8 cm]{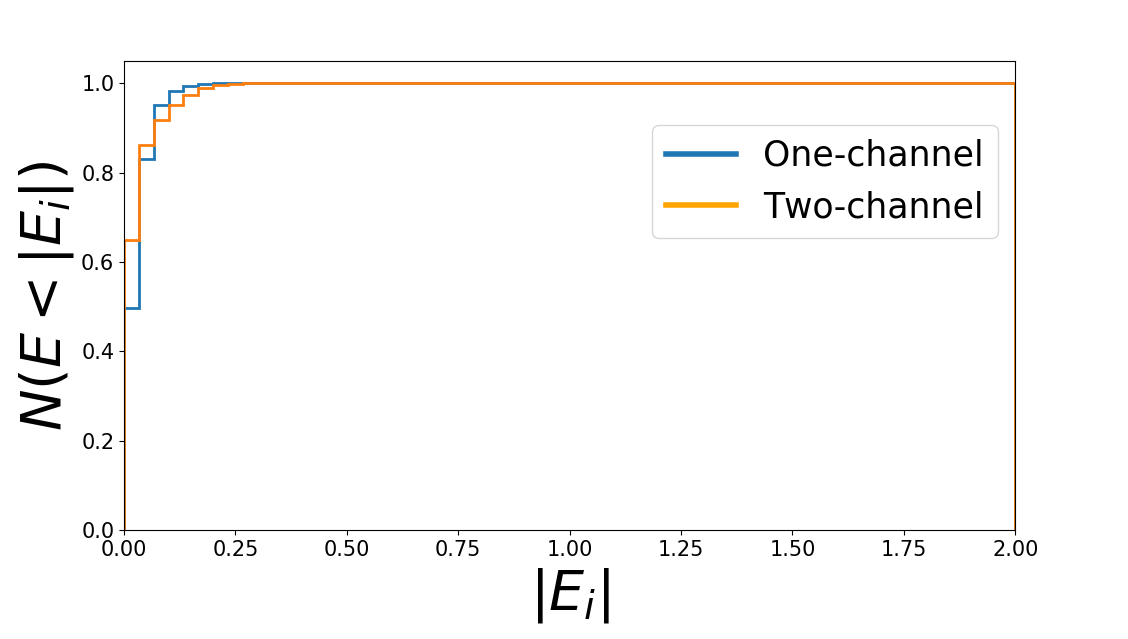}
    \caption{Cumulative error distribution graph for dark matter mass of elliptical galaxies in terms of physical units for one-channel(blue)  and  two-channel  data  (orange).}
    \label{Cumulative Error mergers}
\end{figure}

\subsection{CNN prediction for combined data (Disk+elliptical galaxies)}

\begin{figure}
    \centering
    \includegraphics[width=8 cm]{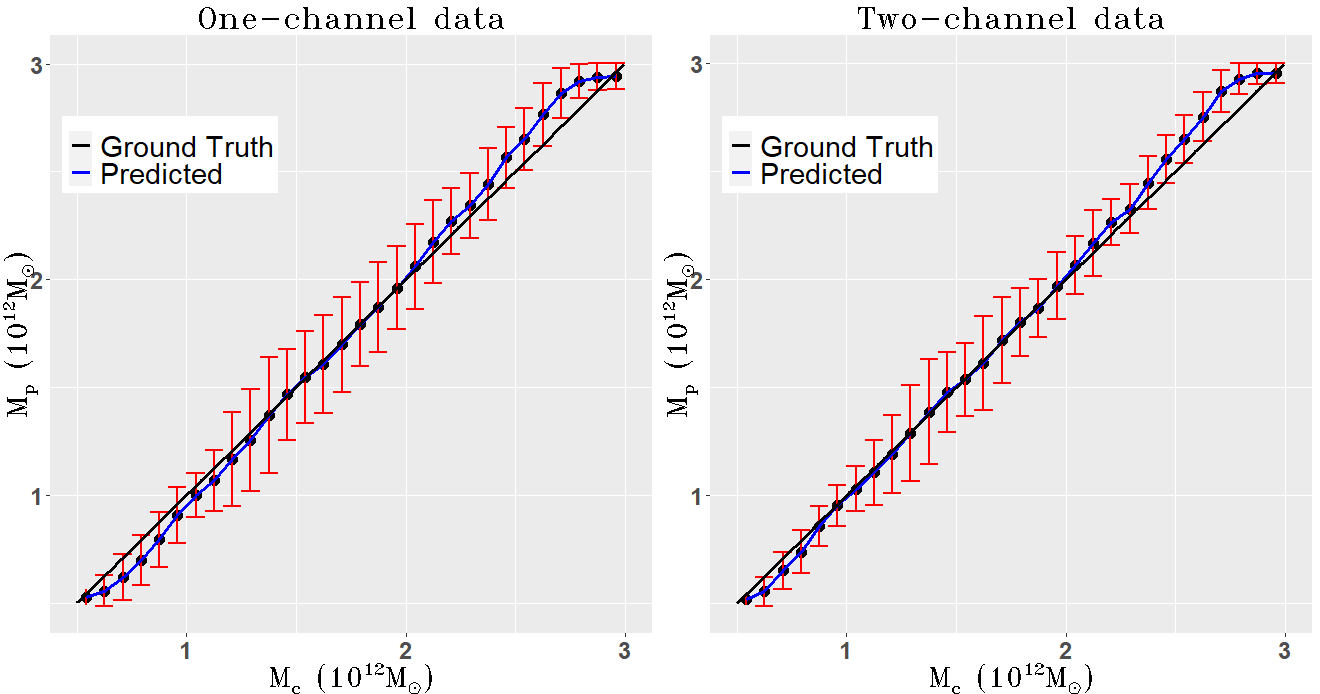}
    \caption{Predicted vs correct dark matter mass within disk and elliptical galaxies for one-channel (left) and two-channel data (right).}
    \label{pre vs cor disk+merger}
\end{figure}

\begin{figure}
    \centering
    \includegraphics[width=8 cm]{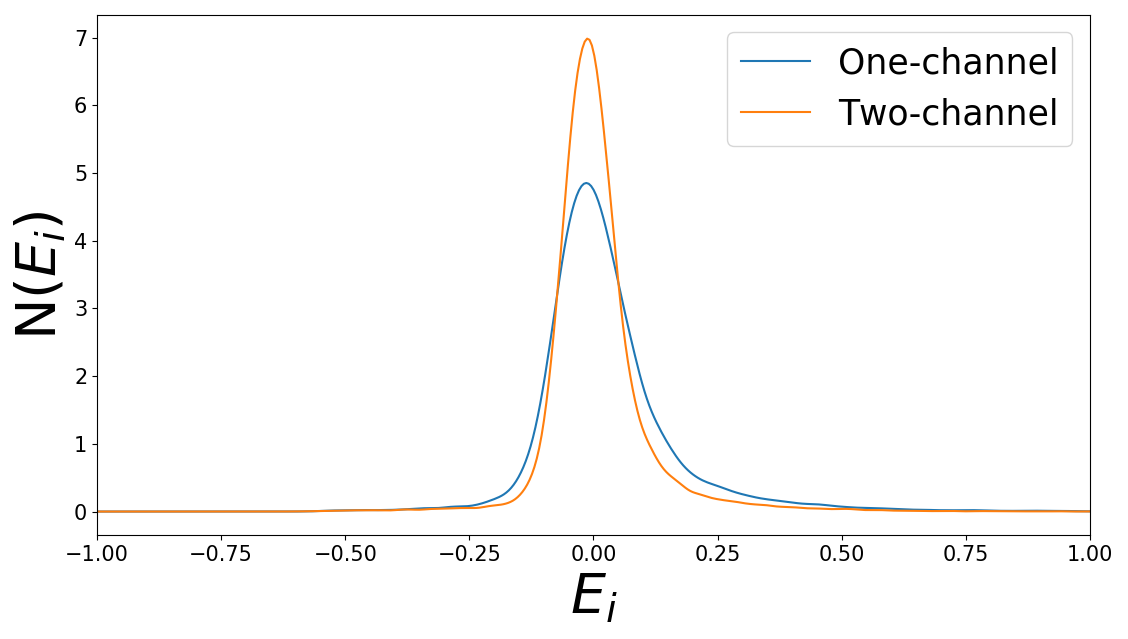}
    \caption{Distribution of errors $E_i$ in the mass estimation of dark matter halos within disk and elliptical galaxies: one-channel (blue) and two-channel (orange).}
    \label{new errors disk+merger}
\end{figure}

We train our CNN using  160,000 of 200,000 images from GCSs in disk and
elliptical galaxies (80\% of data) and then test the CNN by the remaining 40,000 images
(20\%). We do this test in order to confirm that our CNN can accurately predict the 
total masses of galaxies for different types of galaxies.
Figs. \ref{pre vs cor disk+merger} and \ref{new errors disk+merger} demonstrates that
(i) the combination of images from two different types of galaxies
do not influence the prediction accuracy in the CNN-based mass estimation,
(ii) the prediction accuracy does not depend on the mass of galaxies
(i.e., smaller dispersion in each mass bin), (iii) the CNN can predict
the total mass of galaxies with less than 25\% errors for most images,
and (iv) two-channel CNN model can more accurately predict the total galaxy masses than one-channel CNN model.

\begin{figure}
    \centering
    \includegraphics[width=8 cm]{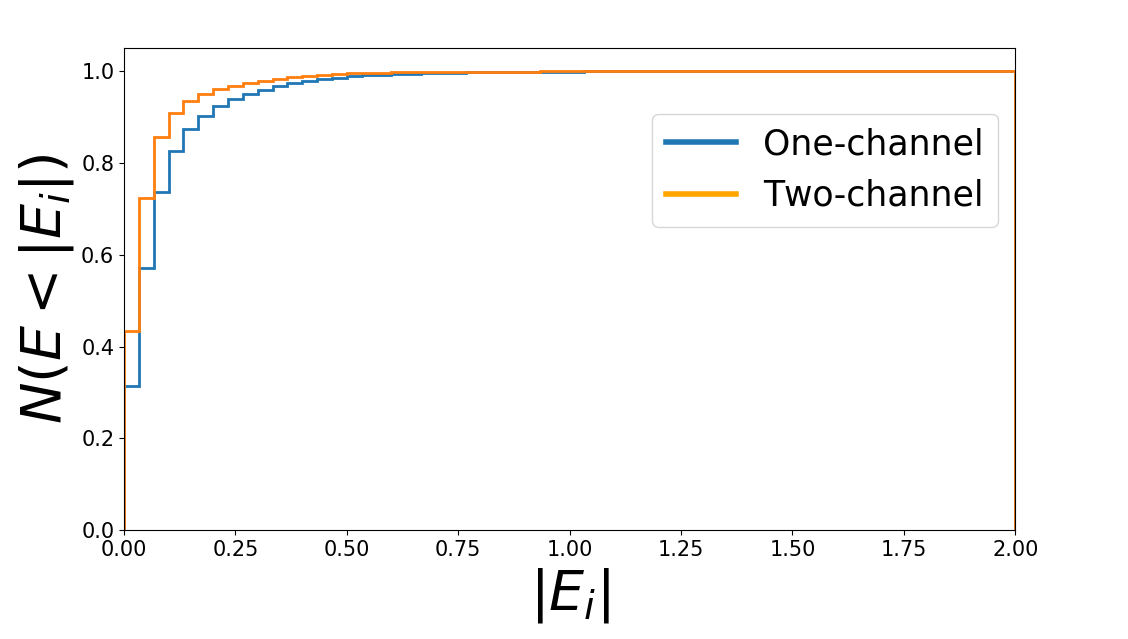}
    \caption{Cumulative error distribution graph for dark matter mass of disk and elliptical galaxies in terms of physical units for one-channel(blue)  and  two-channel data (orange).}
    \label{CED disk+mergers}
\end{figure}

Table \ref{Table 2} shows that
the accuracies for one-channel and two-channel dataset for this
combined model are 0.946 and 0.959 respectively. 
These accuracies are higher than the accuracies from disk model but 
less than merger models. 
Table \ref{MAE & RMSE} shows that 
 RMSEs for one-channel and two-channel data  are 0.733 and 0.595 respectively, 
and MAEs for one-channel and two-channel data are 0.501 and 0.367 
respectively. This better performance can be clearly seen in  
 Fig. \ref{new errors disk+merger} 
and in  Fig. \ref{CED disk+mergers}. 
In Fig. \ref{CED disk+mergers}, our CNN model with 
two-channel data (orange curve) shows around 
45\% of the all images
have less than 5\% of errors. On the other hand, the CNN
model with one-channel data (blue curve) 
shows that about 30\% of the images have less 
than 5\% of errors: the better performance of the two-channel CNN
is confirmed again. 

\subsection{CNN prediction for completely unknown data}
\begin{figure}
    \centering
    \includegraphics[width=8 cm]{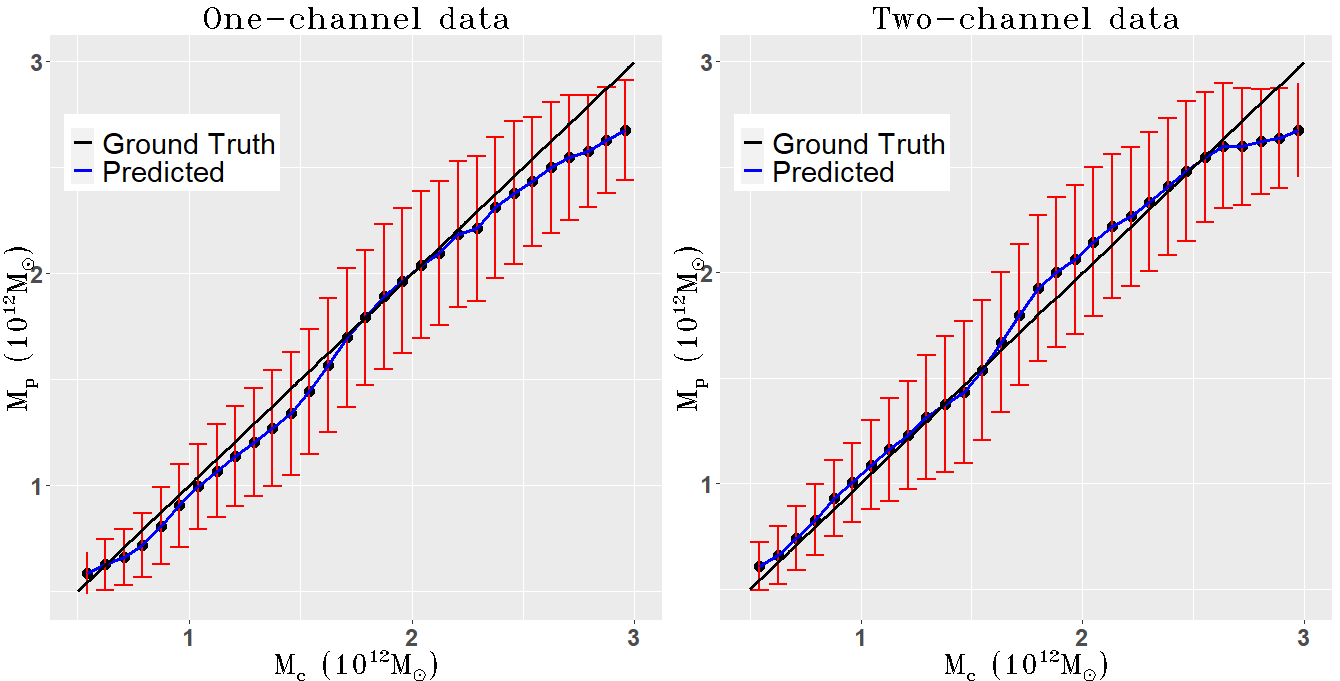}
    \caption{Predicted  vs  correct  dark  matter  mass  within completely unknown galaxies for one-channel (left) and two-channel data (right).}
    \label{pre vs cor disk+merger unknown}
\end{figure}

\begin{figure}
    \centering
    \includegraphics[width=8 cm]{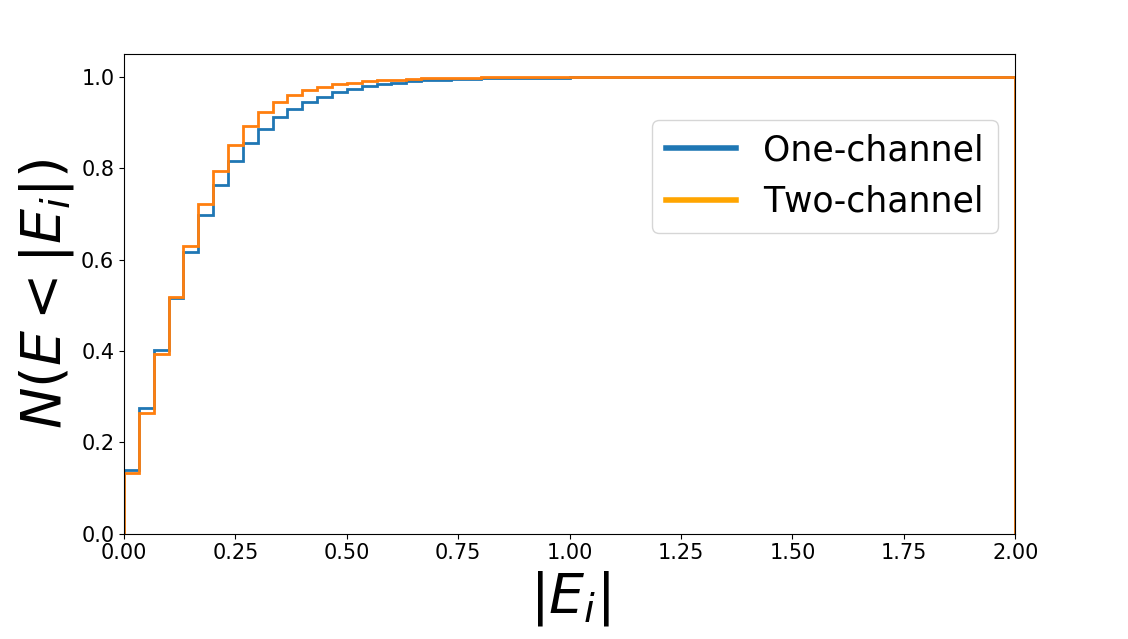}
    \caption{Distribution of errors $E_i$ in the mass estimation of dark matter halos within completely unknown galaxies: one-channel (blue) and two-channel (orange).}
    \label{errors disk+merger unknown}
\end{figure}

\begin{figure}
    \centering
    \includegraphics[width=8 cm]{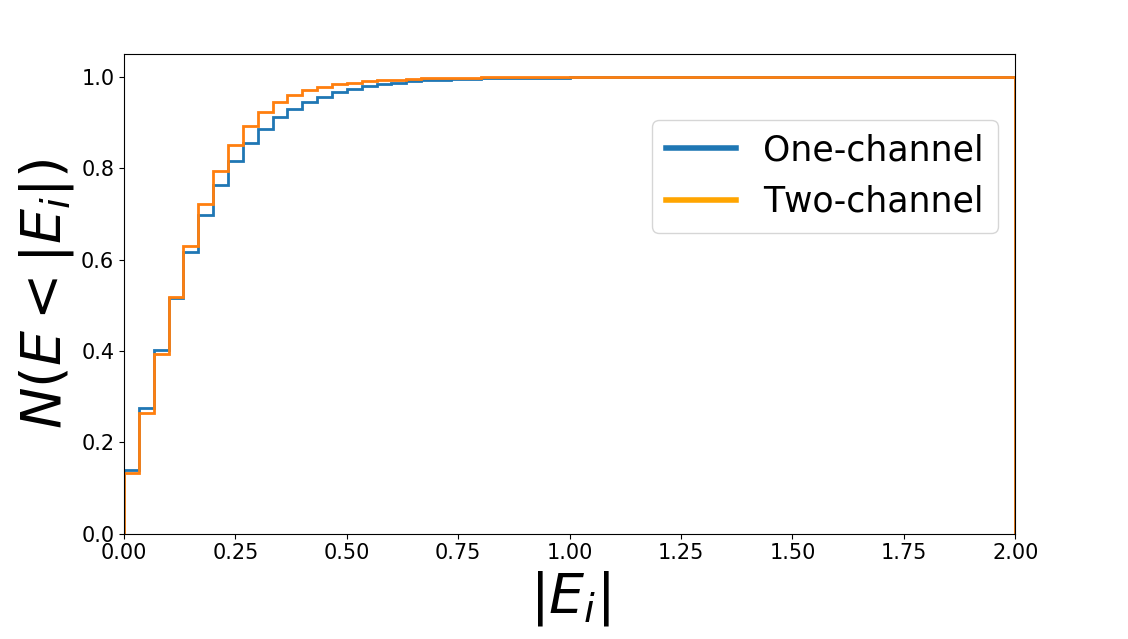}
    \caption{Cumulative error distribution graph for dark matter mass of completely unknown galaxies in terms of physical units for one-channel(blue) and two-channel data (orange).}
    \label{CED disk+merger unknown}
\end{figure}

Here we apply our CNN trained by the combined dataset (disk + elliptical galaxy models) to completely unknown dataset that is unseen by the proposed CNN model. 
This is a more stringent test for our proposed CNN model because these structures and kinematics
of the simulated GCSs in the unknown dataset can be significantly different from
those used for training our model.
The results for this are described in  
Figs, \ref{pre vs cor disk+merger unknown}, \ref{errors disk+merger unknown} and \ref{CED disk+merger unknown}. 
Fig. \ref{pre vs cor disk+merger unknown} confirms
that our CNN performs pretty well even if we apply the CNN to
an entirely unknown dataset. This is a very promising result, which implies that
our CNN will be able to be applied to real observational dataset that is not fully
covered by the present simulations.
Better performance in the two-channel CNN prediction can be clearly seen in  
Fig. \ref{errors disk+merger unknown} 
and
Fig. \ref{CED disk+merger unknown}.

 \subsection{Predictions from a CNN with four convolutional layers}
The results from our original CNN with only two convolutional layers are described
in the preceding subsections.
Here we describe the results from our other CNN with
two additional convolutional layers that are designed
to improve the prediction accuracy in the mass estimation of galaxies from GCS
kinematic
(see the CNN architecture in Fig. \ref{CNN Architecture}).
 Table \ref{Table 2} describes that the prediction accuracies
for this CNN are 0.976 and 0.978 for one-channel and two-channel data 
respectively. Table \ref{MAE & RMSE} shows that
MAEs for one-channel and two-channel data are 0.288 and 0.275, 
and RMSEs are 0.539 and 0.51 for one-channel and two-channel data respectively. 
It is clear that adding two layers in our 
CNN architecture results in an improvement of accuracy both for one-channel and 
two-channel dataset.
However, we confirm that CNNs with more convolutional layers (e.g., 6 of them) cannot further improve the accuracy of prediction.

\section{Discussion}

\subsection{Application of our proposed CNN model to real data}

%%%%% TABLE4
\begin{table*}
\centering
\begin{minipage}{180mm}
\caption{Summary of the total mass of NGC 3115 estimated by proposed CNN model trained by simulated data only. T1-T4 models assumes different velocities and velocity dispersions (2nd column) for meshes having no original data. The observational estimated mass of NGC3115 is $5.7 \pm 0.6 \times 10^{11} {\rm M}_{\odot}$.}\label{Predicted Masses for real data}
\begin{tabular}{lllllllll}
Name
& $V$ and $\sigma$ for meshes with no data
& Predicted Masses (One-channel)
& Predicted Masses (Two-channel)
\\ 
T1 & 0 km s$^{-1}$ & $4.5 \times 10^{11} {\rm M}_{\odot}$ & $5.1 \times 10^{11} {\rm M}_{\odot}$  \\
T2 & 50 km s$^{-1}$ & $4.2 \times 10^{11} {\rm M}_{\odot}$  & $6.0 \times 10^{11} {\rm M}_{\odot}$   \\
T3 & 100 km s$^{-1}$ & $4.5 \times 10^{11} {\rm M}_{\odot}$ & $7.5 \times 10^{11} {\rm M}_{\odot}$  \\
T4 & Interpolated & $6.4 \times 10^{11} {\rm M}_{\odot}$ & $7.3 \times 10^{11} {\rm M}_{\odot}$ \\
\end{tabular}
\end{minipage}
\end{table*}

%%%%% Original 2D Map for Real Data set
\begin{figure}
    \centering
    \includegraphics[width=8 cm]{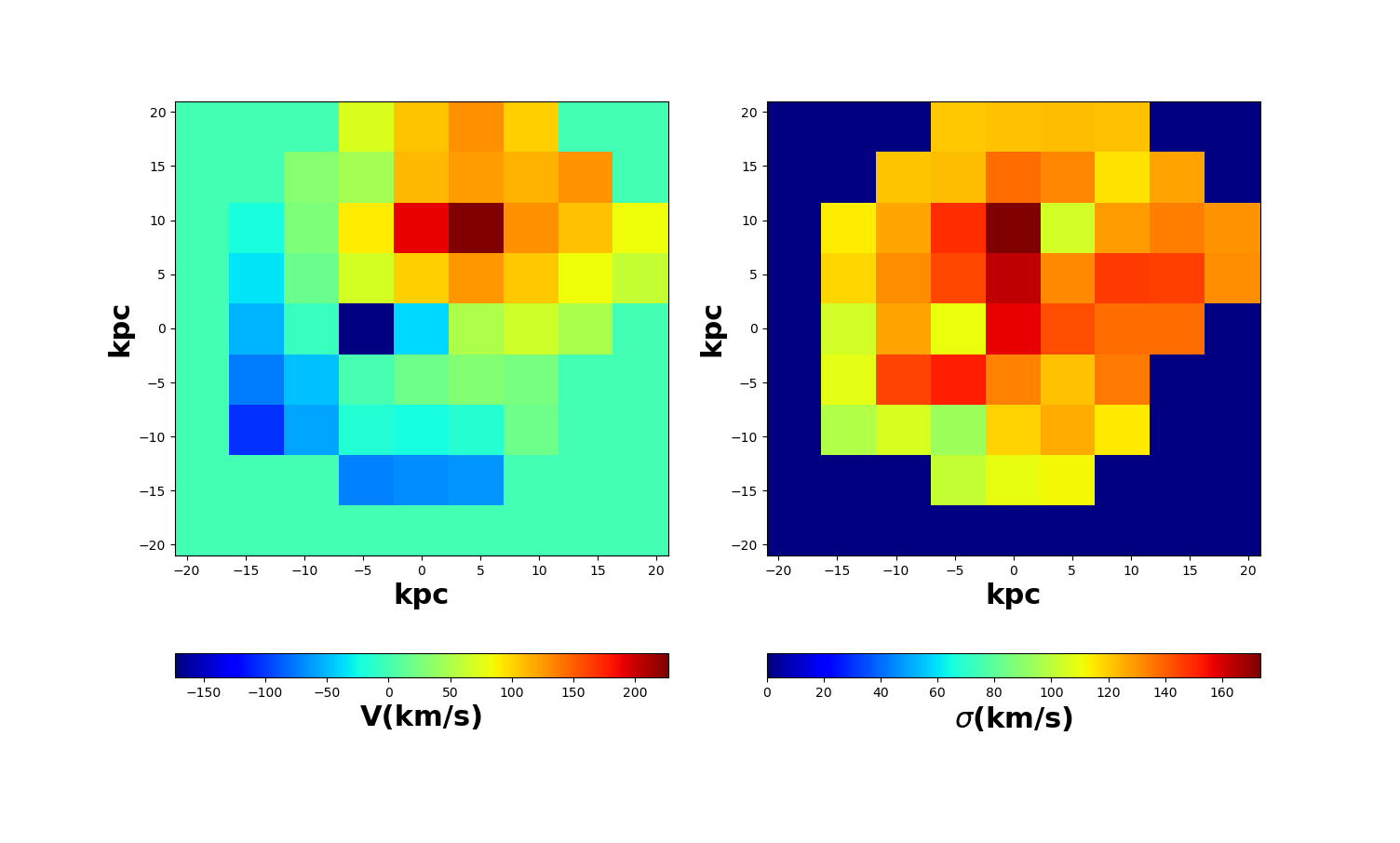}
    \caption{2D maps of line-of-velocity ($V$, left) and  velocity dispersion ($\sigma$, right) in an early-type galaxy NGC 3115. The maps are generated from the original dataset (D20), which has no data in the outer meshes. 0 km/s is allocated for all of these meshes in the two maps. %We discuss how the predicted total mass of this galaxy depends on the assumed velocity and velocity dispersion in these meshes with no original data.
 }
    \label{Original 2D Map for Real dataset}
\end{figure}

Although we have demonstrated that our proposed CNN model can predict the total masses
of galaxies including their dark matter halos quite accurately, we have not
discussed whether the model can predict the galaxy masses accurately when
they are applied to real data. It is observationally difficult to
estimate the total masses of galaxies using GCSs (e.g., Alabi et al. 2016; A16). However, it is possible for the present study to compare between
(i) the observed total masses of galaxies from
other methods using GCSs and (ii) those by the proposed new method.
Observational studies did not investigate the details of 
the 2D kinematics of GCSs in
E/S0 galaxies and thereby make them publicly available. Therefore, we need
to generate such 2D maps by ourselves in the present study. Therefore,
we focus on one object only which is NGC 3115
\citep{Dolfi2020}; D20. One of
the authors of this research article is involved in estimating the 2D kinematics
maps of the GCS, therefore we obtained the data for analysis for this study.
For objects other than NGC 3115, it is possible for us to estimate
the total masses of galaxies from their GCSs using proposed model in future studies
because it is time-consuming to collect all of the required 
radial velocities of their GCs.

\begin{figure}
    \centering
    \includegraphics[width=8 cm]{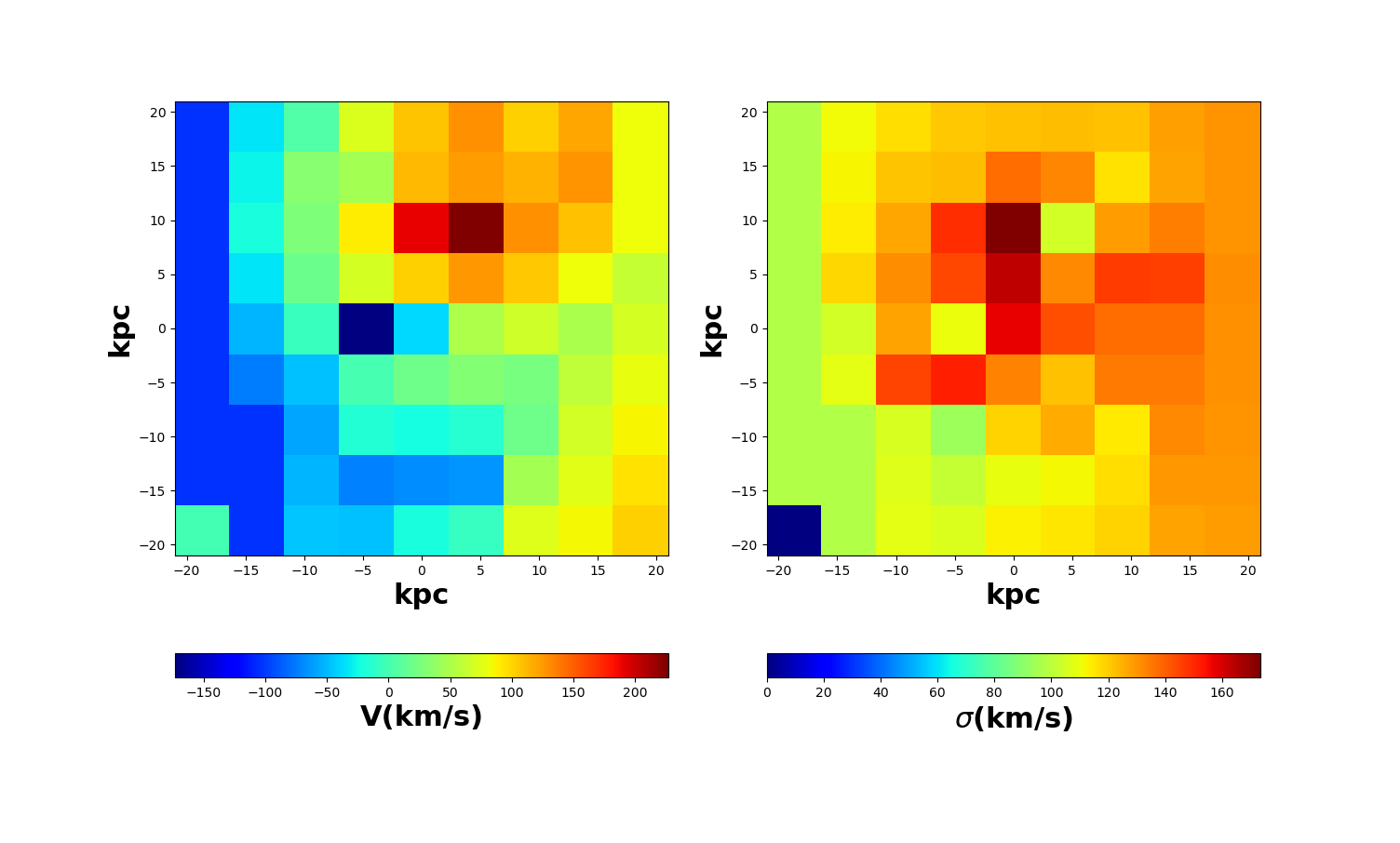}
    \caption{The same as Fig. 17 but for different $V$ and $\sigma$ for meshes having no original data. The interpolated
values are used for these meshes.
}
    \label{2D Map for Interpolated values}
\end{figure}

First, we produce the 2D velocity and dispersion maps of the GCs in NGC 3115
in order to input the 2D maps into our CNN model trained by simulation dataset.
Fig. \ref{Original 2D Map for Real dataset} shows the 2D maps generated from the original data set (D20),
which has no data in the outer part  of the galaxy ($\sim 30$ pixels)
due to no observations (i.e., no GCs)
for these pixels. This lack of data in the original image can prevent us
from estimating the total mass of the galaxy. However, we can still
apply our CNN model to this image by assuming velocities and velocity dispersion in these pixels. Here we generate four test images from the original 2D maps,
T1, T2, T3, and T4, as follows. For T1, T2, and T3, the pixels with no original
data have 0, 50, and 100 km s$^{-1}$, respectively, so that we can discuss
how the assumed values in the pixels influence the mass estimation of galaxies.
T4 is assumed to have new values interpolated from other pixels for the pixels having no original data. 
We use the linear interpolation method by ``Pandas" python library in order
to generate new 2D maps in T4 in which all pixels have velocity
and velocity dispersion values. Fig. \ref{2D Map for Interpolated values} show the new maps generated in T4.

%%%%%%% 2D Map with Interpolated values

Table \ref{Predicted Masses for real data} presents the total masses of NGC 3115 estimated by our proposed CNN model for T1, T2, T3 and T4 testing models %, and this is true for observations by A16. 
Table \ref{Predicted Masses for real data} also presents the predicted galaxy masses from one-channel
and two-channel by the proposed CNN model.
It is clear from the results presented in Table \ref{Predicted Masses for real data} that the T1 model underestimates the total mass of this galaxy: the predicted mass is $4.5 \times 10^{11} {\rm M}_{\odot}$ that is significantly
smaller than the observationally estimated mass \textbf{$5.7 \pm 0.6 \times 10^{11} {\rm M}_{\odot}$}. This is mainly
due to the reason that the velocity dispersion map in T1 has an unrealistically
steeper radial gradient in the outer parts, which can end up with the lower total mass. This underestimation can also be observed from the two-channel results for the T1 model but the total mass
predicted by the proposed CNN model is closer to the observed one.
Such an underestimation of galactic masses can also be observed in T2 and T3 models for
one-channel. However, the estimated mass by two-channel
is $6.0 \times 10^{11} {\rm M}_{\odot}$ and $7.5 \times 10^{11} {\rm M}_{\odot}$,
for T2 and T3 respectively. 
The total galactic mass predicted by two-channel CNN in T4 model
is $7.3 \times 10^{11} {\rm M}_{\odot}$. Hence for one channel data T4 (Interpolated) model performs better as it is closer to the observed value. However, for two-channel, T1 and T2 models perform better. Overall, we conclude that all models performs very close to the observed mass. This
strongly suggests that if the 2D kinematics maps of GCs in a galaxy
can be properly
generated through a linear interpolation method,  the total mass
can be accurately predicted by our CNN model.

Since we have so far applied our CNN model to only one GCS, we cannot make a robust
conclusion as to whether our proposed can be a new, practical
way to estimate the total
masses of galaxies precisely (like other existing methods).
Furthermore, it should be noted here that the ``observed mass'' of a galaxy
might not be 
the real/true mass of the galaxy: it is just an estimated mass from other
existing methods that are used frequently in other studies.
Therefore, the above-mentioned comparison between the estimation by our proposed CNN model
and the ``observed mass'' may not be the real comparison between
the CNN-based mass estimation and the observed mass.
Accordingly, we plan the pursue our research on following two areas.
First, we plan to apply our proposed CNN model to larger number
of the observed GCS kinematics and thereby compare the 
estimated masses with those from other methods: A16 already provided the
number of galaxies with known total masses.
Second, we plan to apply our proposed CNN model and other methods to the same simulation
dataset for which the total masses of galaxies are known. This second
investigation will allow us to discuss whether our proposed model or other methods
perform better to estimate the total masses of galaxies from their GCS kinematics.

\subsection{Advantages and disadvantages of the new method}

We have demonstrated that the new method based on deep learning can predict
the total masses of galaxies pretty well, though the number of images
used for training CNNs is not huge \textbf{(``in order of $10^6$ samples'').} The mean  accuracy of
the prediction can reach 98\% (i.e., the predicted masses deviate from the
true values by only 2\%), which implies that the new method will be able to
be applied to the kinematics of the observed GCSs. Furthermore, the better
prediction accuracy for two-channel data ($V$ and $\sigma$) implies that 
global rotation of GCSs needs to be incorporated into the mass estimation
models for more accurate predictions of total masses in galaxies. 
This also implies that the total masses of galaxies with GCSs with 
higher amplitudes of global rotation, e.g., NGC 3115 \citep[]{Dolfi2020}
can be better estimated by the present new method than others that only
consider the radial profiles of $\sigma$ for GCSs.
We suggest that if
the 2D kinematics of GCSs predicted from cosmological simulations
of GC formation, \citep[e.g.,][]{Pfeffer2017}  can be
used to generate the synthesized 2D maps of GCS kinematics that can be input
to CNNs,  an even better mass estimator based on deep learning can be 
developed, because the predicted kinematics can be more diverse and more realistic
than the synthesized images adopted in the present study.

The new method, however, has the following possible problems, which need to be 
solved in our future studies.
First, the total number of GCs in a galaxy should be at least several tens to
$\sim 100$
so that the 2D maps of $V$ and $\sigma$
with enough spatial resolutions (e.g., $10 \times 10$ pixels)
can be constructed.
This means that the new method cannot be properly applied to GCSs of less
luminous galaxies like NGC 4564 with $N_{\rm gc}=27$ \citep[]{Alabi2016}.
If there is no GC in a significant number of pixels for less luminous
galaxies, a large smoothing length need to be applied to the 2D maps
so that smooth $V$ and $\sigma$ fields can be derived.
However, such a large smoothing length would cause a significant 
reduction of the  prediction accuracy in the mass estimation.
Second, it it not so clear how sensitive the new mass estimation method is
to the presence of substructures in GCSs. In the present disk and elliptical
galaxy models with GCSs, the synthesized images do not show substructures clearly
whereas some of the observed GCSs show substructures \citep[]{Alabi2016}.
This means that the training dataset should contain GCSs with substructures
for more accurate predictions of total galactic masses 
in our future study.

\begin{figure*}
    \centering
    \includegraphics[width=18 cm]{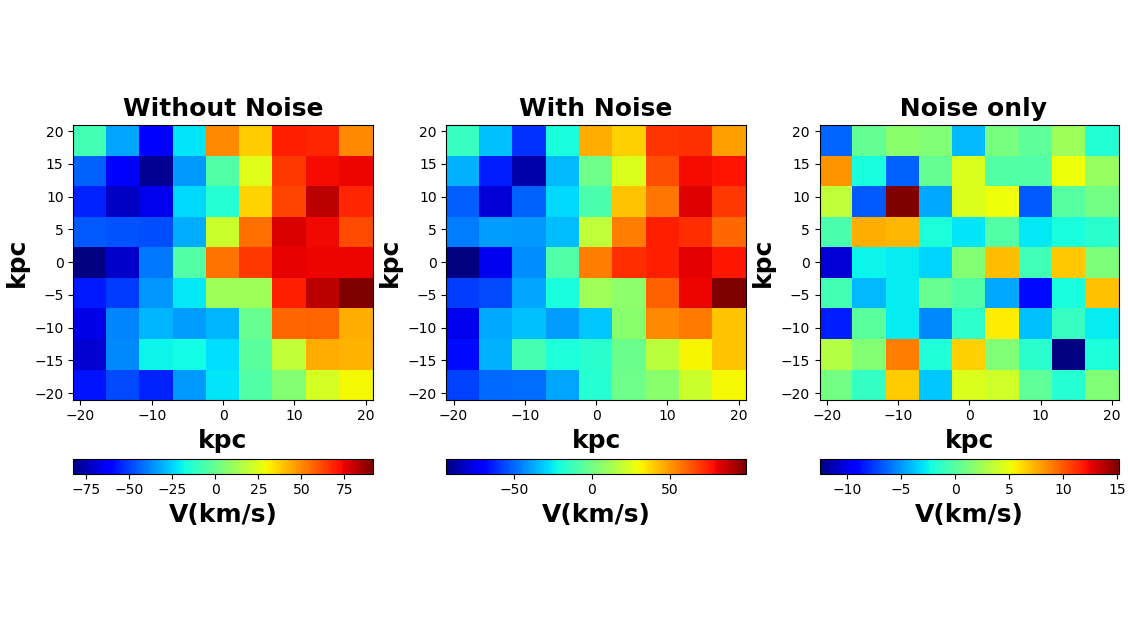}
    \caption{2D map of $V$ without (left) and with noise (middle) for the GCS of a isolated disk galaxy. Noise with the mean of 0 and the standard deviation of 0.04 (in simulation units, corresponding to 5 km/s) is given to each mesh point in the middle panel. The right panel shows noise map.}
    \label{noise}
\end{figure*}

Third, observational errors in the spectroscopic
estimation of line-of-sight-velocities ($V$) for individual GCs can possibly
introduce noise in the 2D maps of GCSs. If CNNs are trained by 2D images
with such noise, then the prediction accuracy will be reduced significantly too.
In order to investigate this issue more quantitatively,  
we add Gaussian noise manually to the image data and thereby
investigate the performance of our CNNs. We add noise with mean zero 
and standard deviation of 0.04 to each pixel in this investigation. This error correspond
to spectroscopic errors of $\sim 5$ km s$^{-1}$ for $V$ and $\sigma$. The $\sim 5$ km s$^{-1}$ error in each mesh corresponds to the observationally possible errors for radial velocities of GCs \citep[]{Dolfi2020}. Therefore our adoption of this value is consistent with observations of GCSs. Since there can be a number of GCs in one pixel, it is 
not straightforward to convert observational errors for one GC into
noise (errors) for the estimated $V$ and $\sigma$ for one pixel,
this introduction of noise enables us to discuss this important issues
in a quantitative manner.
Fig. \ref{noise} represents the 2D map of the image without noise, with noise,
and noise only. It is found
that after adding noise to images, our model accuracy significantly 
drops from 0.976 to 0.809 for two-channel combined data from 
disk and elliptical galaxies (disk+merger). Moreover, RMSEs for one-channel and two-channel data for images with noise are 3.234 and 3.408, 
respectively, and MAEs for one-channel and two-channel data are 2.722 and 2.835, 
respectively. This means that
 our model still performs well on the dataset containing noise,
though the prediction accuracy would become worse for noise of 
more than $10$ km s$^{-1}$.

\subsection{Toward the application of the new method to observations}

Although we have developed the new mass estimation method based on deep learning,
this work can be regarded as a  proof-of-concept.
In order to apply the new method to real observations, we need to (i)  solve
the above-mentioned three possible problems related to
the implementation of the new method and  (ii) significantly  increase the number of 
synthesized images of GCSs not only from constrained simulations (like the
present study) but also from cosmological ones.
The above second point is crucial for our future work, because the simulated
kinematics of GCSs from isolated disk galaxies and major mergers is quite limited
and possibly could not represent a full range of the observed kinematics
of GCSs with diverse rotation amplitudes and anisotropy parameters 
(i.e., tangentially or radially anisotropic).

The formation and evolution of GCSs in galaxies with different masses and types
can involve various physical processes of galaxy formation
(not just major merging and secular disk
evolution modeled in this study).
Therefore,
it is ideal for our future study to generate the synthesized 2D kinematics of
GCSs in galaxies from cosmological simulations that can model various
physical processes of galaxy formation from the early universe.
Since previous and recent cosmological simulations of GCS formation in galaxies
have already predicted the kinematics of GCSs, \citep[e.g.,][]{Bekki2008,Pfeffer2017}, 
it will be feasible for our future study to generate a large number of
2D kinematics maps of GCSs from these simulations, which will be input
to CNNs for better prediction accuracies.

\section{Conclusions}

We have developed a new method by which we can estimate the total masses of 
galaxies including their dark matter halos 
from the kinematics of their GCSs. 
This new method is based on deep learning in which
a large number of synthesized 2D maps of kinematics of GCSs from numerical
simulations of galaxies with known total masses are used to predict
the total masses of galaxies. The new mass estimation method has 
involved the following two steps.
In the first step of the  method, 
we have trained a convolutional neutral network (CNN) 
using a large number (more than 100,000)
of synthesized 2D kinematics maps (``images'')  of GCSs from 
simulated disk and elliptical galaxies with known total masses.
Accordingly, the input data to a CNN is the large number pairs of (i)
2D kinematics maps of GCS and (ii) true (known) total masses of galaxies
including dark matter.

In this supervised learning phase,
either only 2D maps of velocity dispersions ($\sigma$) of GCSs (``one-channel'')
or 2D maps of both $\sigma$ and $V$ (line-of-sight-velocities)  of GCSs
(``two-channel'')
are used so that we can determine whether one-channel or two-channel
prediction can be more accurate.
This large number of images can cover a wide range of GCS kinematics
with various degrees of global rotation and anisotropy parameters,
though the present study possibly
would not  cover a full range of the  observed GCS kinematics.
In the second step of the  method, 
we have applied the CNN to a completely unknown dataset to investigate
whether or not the CNN can accurately predict the true masses of galaxies.

For disk galaxies, we have used  100,000 images of GCSs kinematics
from different disk galaxies with different degrees of rotational kinematics in
GCSs.
We have used  80,000 images of GCSs kinematics for elliptical galaxies,
and also combined these images from the two types of galaxies.
In order to investigate how observational errors (represented as noise
in pixels) can influence the prediction accuracy of the new mass
estimation method,
we have added Gaussian noise to the images 
from the combined dataset (disk+merger) with mean $0$ and standard deviation 
of 5 km s$^{-1}$. 
%We have demonstrated that the new CNN-based technique can 
%accurately estimate the total masses of galaxies, 
%which is highly competitive with standard analysis techniques. 
Furthermore
we have verified the prediction accuracy of our CNN using RMSEs and MAEs for each 
model. We have compared between different architectures of CNNs in terms of
prediction accuracies and shown the results from the best CNN architecture in the 
present study.
The principal results are as follows: \\

(1) Our new model based on our original CNN can predict the total masses of disk galaxies
dominated by  dark matter pretty accurately. 
The overall accuracy of the model trained by only 2D maps of $\sigma$ of GCSs
(one-channel data) is 93.7\%. However,  
the accuracy of the model trained by 2D maps of $V$ and $\sigma$
(two-channel data) is improved by about 3\%.
We have observed MAE of one-channel data is 0.571 and for two-channel data, it is 0.404,
which clearly demonstrate a very  high prediction accuracy.
Moreover, RMSEs for one-channel and two-channel data are 0.833 and 0.707, 
respectively. This implies that the model with two-channel data works better 
than one-channel data, as two-channel data has a smaller MAE. This is mainly
because the rotation energy of GCSs can be properly considered in the mass
estimation of galaxies including dark matter:
this contribution from global rotation should be considered in the
mass estimation.

(2) Our model can also accurately 
predict the total masses of elliptical galaxies including
their dark matter. 
The observed accuracies for the model trained by
one-channel and two-channel data are 97.2\%  and 98\%, respectively. 
These accuracies are slightly better than our results 
for disk galaxies. The outcomes for RMSEs for elliptical galaxies 
with one-channel and two-channel data are 0.357 and 0.239, 
respectively. Similarly, the value of MAEs for the model with 
one-channel and two-channel data are 0.265 and 0.188, respectively. 
In this case, 
it is again observed that two-channel data outperforms 
one-channel data as it has higher accuracy and lower RMSE and MAE values.

(3) We have combined the data from both disk and
elliptical galaxies and thereby trained CNN with a 
large number of images (200,000). The overall accuracies 
for the combined dataset for one-channel and two-channel 
data are 94.6\% and 95.9\% respectively. 
Moreover, results for RMSEs for one-channel and 
two-channel data are 0.733 and 0.595, respectively.
It is also observed that MAEs for combined dataset 
(disk+elliptical) for model with one-channel and 
two-channel data are 0.501 and 0.367, 
respectively. This shows that two-channel data
again outperforms one-channel data for this.
These results imply that the new method can be applied to different types
of galaxies with GCSs.

(4) We have also  used a CNN with two more convolutional layers in order
to improve accuracy. We have found that 
our accuracy for the model with combined data (disk+ elliptical galaxies)
is increased by 3\% for one-channel (97.6\%) and 2\% for two-channel (97.8\%).
The RMSEs for the new architecture are 0.539 and 0.51 for one-channel and 
two-channel, respectively. 
Moreover, MAEs for the model with one-channel and two-channel data are 
0.288 and 0.275, respectively. 
Hence, it is observed that the accuracies of the model 
(one-channel and two-channel) are significantly improved with the addition of 
two more convolutional layers in the existing architecture.

(5) In the above tests, we have used 80\% of the original data
for training CNNs and used the 20\% for testing.
This means that  the testing dataset can be similar
to the training data to some extent, because the two dataset generated
from GCSs in the same simulated galaxies. 
In order to do a more stringent test for the new CNNs,
we have applied our CNN to completely unknown data that is not at all even
for training.
We have confirmed that our CNN
can accurately predict the galaxy masses for this completely unknown data.

(6) We have found that  CNN
trained by  200,000 images containing 
Gaussian noise 
predicts
the total masses of
with the accuracy of 80.9\%, which implies that the prediction accuracy is still good,
though it needs to be improved through the revision of the CNN architecture
in our future studies.

(7) We applied our CNN model trained by simulation data to real data of the GCS in NGC 3115 (an
early-type galaxy) in order to compare the estimated total mass by our model with an existing methods (A16). Although it is not observationally clear whether the total mass
estimated by A16 is really a true value, the total mass predicted by our CNN model is similar to A16. In particular, our proposed model trained by two-channel dataset shows a better consistency
with the results of A16. We have a plan to apply our model to other GCSs in other E/S0 galaxies
in our future studies.

(8) Although we have successfully demonstrated that the new method can accurately
predict the total masses of galaxies, the simulated GCS kinematics can be very
limited (not so diverse as real observations). 
Therefore, in order to apply the method to real observations,
we need to dramatically increase the number of 2D kinematics maps
of GCSs that cover a large range of the possible GCS kinematics in real
galaxies. Thus we will use the predictions from cosmological
simulations of galaxy formation with GCS models in order to train our
CNNs in our future studies.

\section{Acknowledgment}
We are grateful to the reviewers for their constructive and
useful feedback that improved this paper. We appreciate Arianna Dolfi who kindly estimated the 2D velocity and velocity dispersion maps of all the GCs of NGC 3115 and provide us these observations for our study.

\section{Data Availability}

The data underlying this article will be shared on a justified request to the corresponding author.

\bibliography{paper.bib}{}

\begin{thebibliography}{}
\makeatletter
\relax
\def\mn@urlcharsother{\let\do\@makeother \do\$\do\&\do\#\do\^\do\_\do\%\do\~}
\def\mn@doi{\begingroup\mn@urlcharsother \@ifnextchar [ {\mn@doi@}
  {\mn@doi@[]}}
\def\mn@doi@[#1]#2{\def\@tempa{#1}\ifx\@tempa\@empty \href
  {http://dx.doi.org/#2} {doi:#2}\else \href {http://dx.doi.org/#2} {#1}\fi
  \endgroup}
\def\mn@eprint#1#2{\mn@eprint@#1:#2::\@nil}
\def\mn@eprint@arXiv#1{\href {http://arxiv.org/abs/#1} {{\tt arXiv:#1}}}
\def\mn@eprint@dblp#1{\href {http://dblp.uni-trier.de/rec/bibtex/#1.xml}
  {dblp:#1}}
\def\mn@eprint@#1:#2:#3:#4\@nil{\def\@tempa {#1}\def\@tempb {#2}\def\@tempc
  {#3}\ifx \@tempc \@empty \let \@tempc \@tempb \let \@tempb \@tempa \fi \ifx
  \@tempb \@empty \def\@tempb {arXiv}\fi \@ifundefined
  {mn@eprint@\@tempb}{\@tempb:\@tempc}{\expandafter \expandafter \csname
  mn@eprint@\@tempb\endcsname \expandafter{\@tempc}}}

\bibitem[\protect\citeauthoryear{Alabi et~al.,}{Alabi et~al.}{2016}]{Alabi2016}
Alabi A.~B.,  et~al., 2016, Monthly Notices of the Royal Astronomical Society,
  460, 3838

\bibitem[\protect\citeauthoryear{Alabi et~al.,}{Alabi et~al.}{2017}]{Alabi2017}
Alabi A.~B.,  et~al., 2017, Monthly Notices of the Royal Astronomical Society,
  468, 3949

\bibitem[\protect\citeauthoryear{Bahcall \& Tremaine}{Bahcall \&
  Tremaine}{1981}]{Bahcall1981}
Bahcall J.~N.,  Tremaine S.,  1981, The Astrophysical Journal, 244, 805

\bibitem[\protect\citeauthoryear{Beasley \& Trujillo}{Beasley \&
  Trujillo}{2016}]{Beasley2016}
Beasley M.~A.,  Trujillo I.,  2016, The Astrophysical Journal, 830, 23

\bibitem[\protect\citeauthoryear{Bekki}{Bekki}{2013}]{Bekki2013}
Bekki K.,  2013, Monthly Notices of the Royal Astronomical Society, 432,
  2298–2323

\bibitem[\protect\citeauthoryear{{Bekki}, {Couch}, {Forbes}  \&
  {Beasley}}{{Bekki} et~al.}{2005}]{Bekki2005}
{Bekki} K.,  {Couch} W.~J.,  {Forbes} D.~A.,   {Beasley} M.~A.,  2005,
  Highlights of Astronomy, 13, 191

\bibitem[\protect\citeauthoryear{Bekki, Yahagi, Nagashima  \& Forbes}{Bekki
  et~al.}{2008}]{Bekki2008}
Bekki K.,  Yahagi H.,  Nagashima M.,   Forbes D.~A.,  2008, Monthly Notices of
  the Royal Astronomical Society, 387, 1131

\bibitem[\protect\citeauthoryear{Bekki, Diaz  \& Stanley}{Bekki
  et~al.}{2019}]{Bekki2019}
Bekki K.,  Diaz J.,   Stanley N.,  2019, Astronomy and Computing, 28, 100286

\bibitem[\protect\citeauthoryear{Boylan-Kolchin}{Boylan-Kolchin}{2018}]{Boylan-Kolchin2018}
Boylan-Kolchin M.,  2018, Monthly Notices of the Royal Astronomical Society,
  479, 332

\bibitem[\protect\citeauthoryear{Brodie \& Strader}{Brodie \&
  Strader}{2006}]{Brodie2006}
Brodie J.~P.,  Strader J.,  2006, Annual Review of Astronomy and Astrophysics,
  44, 193

\bibitem[\protect\citeauthoryear{Brodie et~al.,}{Brodie
  et~al.}{2014}]{Brodie2014}
Brodie J.~P.,  et~al., 2014, The Astrophysical Journal, 796, 52

\bibitem[\protect\citeauthoryear{{Cavanagh} \& {Bekki}}{{Cavanagh} \&
  {Bekki}}{2020}]{Cavanagh2020}
{Cavanagh} M.,  {Bekki} K.,  2020, arXiv e-prints, p. arXiv:2006.14847

\bibitem[\protect\citeauthoryear{Diaz, Bekki, Forbes, Couch, Drinkwater  \&
  Deeley}{Diaz et~al.}{2019}]{Diaz2019}
Diaz J.~D.,  Bekki K.,  Forbes D.~A.,  Couch W.~J.,  Drinkwater M.~J.,   Deeley
  S.,  2019, Monthly Notices of the Royal Astronomical Society, 486, 4845

\bibitem[\protect\citeauthoryear{{Dieleman}, {Willett}  \& {Dambre}}{{Dieleman}
  et~al.}{2015}]{Dieleman2015}
{Dieleman} S.,  {Willett} K.~W.,   {Dambre} J.,  2015, Monthly Notices of the
  Royal Astronomical Society, 450, 1441

\bibitem[\protect\citeauthoryear{Dolfi et~al.,}{Dolfi et~al.}{2020}]{Dolfi2020}
Dolfi A.,  et~al., 2020, Monthly Notices of the Royal Astronomical Society,
  495, 1321

\bibitem[\protect\citeauthoryear{{Forbes}, {Read}, {Gieles}  \&
  {Collins}}{{Forbes} et~al.}{2018}]{Forbes2018}
{Forbes} D.~A.,  {Read} J.~I.,  {Gieles} M.,   {Collins} M. L.~M.,  2018,
  Monthly Notices of the Royal Astronomical Society, 481, 5592

\bibitem[\protect\citeauthoryear{G\'{e}ron}{G\'{e}ron}{2017}]{Geron2017}
G\'{e}ron A.,  2017, Hands-on machine learning with Scikit-Learn and TensorFlow
  : concepts, tools, and techniques to build intelligent systems.
O'Reilly Media, Sebastopol, CA

\bibitem[\protect\citeauthoryear{Griffen, Drinkwater, Iliev, Thomas  \&
  Mellema}{Griffen et~al.}{2013}]{Griffen2013}
Griffen B.~F.,  Drinkwater M.~J.,  Iliev I.~T.,  Thomas P.~A.,   Mellema G.,
  2013, Monthly Notices of the Royal Astronomical Society, 431, 3087

\bibitem[\protect\citeauthoryear{{Hariharan}, {Arbeláez}, {Girshick}  \&
  {Malik}}{{Hariharan} et~al.}{2017}]{Hariharan2017}
{Hariharan} B.,  {Arbeláez} P.,  {Girshick} R.,   {Malik} J.,  2017, IEEE
  Transactions on Pattern Analysis and Machine Intelligence, 39, 627

\bibitem[\protect\citeauthoryear{{Hinton}, {Srivastava}, {Krizhevsky},
  {Sutskever}  \& {Salakhutdinov}}{{Hinton} et~al.}{2012}]{Hinton2012}
{Hinton} G.~E.,  {Srivastava} N.,  {Krizhevsky} A.,  {Sutskever} I.,
  {Salakhutdinov} R.~R.,  2012, arXiv e-prints, p. arXiv:1207.0580

\bibitem[\protect\citeauthoryear{Krizhevsky, Sutskever  \& Hinton}{Krizhevsky
  et~al.}{2012}]{Alex2012}
Krizhevsky A.,  Sutskever I.,   Hinton G.~E.,  2012, Communications of the ACM,
  60, 84

\bibitem[\protect\citeauthoryear{Krizhevsky, Sutskever  \& Hinton}{Krizhevsky
  et~al.}{2017}]{Krizhevsky2017}
Krizhevsky A.,  Sutskever I.,   Hinton G.~E.,  2017, Communications of the ACM,
  60, 84–90

\bibitem[\protect\citeauthoryear{Morganti, Gerhard, Coccato, Martinez-Valpuesta
   \& Arnaboldi}{Morganti et~al.}{2013}]{Morganti2013}
Morganti L.,  Gerhard O.,  Coccato L.,  Martinez-Valpuesta I.,   Arnaboldi M.,
  2013, Monthly Notices of the Royal Astronomical Society, 431, 3570

\bibitem[\protect\citeauthoryear{{Navarro}, {Frenk}  \& {White}}{{Navarro}
  et~al.}{1996}]{Novarro1996}
{Navarro} J.~F.,  {Frenk} C.~S.,   {White} S. D.~M.,  1996, The Astrophysical
  Journal, 462, 563

\bibitem[\protect\citeauthoryear{{Neto} et~al.,}{{Neto}
  et~al.}{2007}]{Neto2007}
{Neto} A.~F.,  et~al., 2007, Monthly Notices of the Royal Astronomical Society,
  381, 1450

\bibitem[\protect\citeauthoryear{Peng, Ford  \& Freeman}{Peng
  et~al.}{2004}]{Peng2004}
Peng E.~W.,  Ford H.~C.,   Freeman K.~C.,  2004, The Astrophysical Journal,
  602, 705

\bibitem[\protect\citeauthoryear{Pfeffer, Kruijssen, Crain  \& Bastian}{Pfeffer
  et~al.}{2017}]{Pfeffer2017}
Pfeffer J.,  Kruijssen J. M.~D.,  Crain R.~A.,   Bastian N.,  2017, Monthly
  Notices of the Royal Astronomical Society, 475, 4309

\bibitem[\protect\citeauthoryear{Prole et~al.,}{Prole et~al.}{2019}]{Prole2019}
Prole D.~J.,  et~al., 2019, Monthly Notices of the Royal Astronomical Society,
  484, 4865

\bibitem[\protect\citeauthoryear{{Romanowsky}, {Douglas}, {Arnaboldi},
  {Kuijken}, {Merrifield}, {Napolitano}, {Capaccioli}  \&
  {Freeman}}{{Romanowsky} et~al.}{2003}]{Romanowsky2003}
{Romanowsky} A.~J.,  {Douglas} N.~G.,  {Arnaboldi} M.,  {Kuijken} K.,
  {Merrifield} M.~R.,  {Napolitano} N.~R.,  {Capaccioli} M.,   {Freeman} K.~C.,
   2003, Science, 301, 1696

\bibitem[\protect\citeauthoryear{{Santos}, {Cooray}, {Haiman}, {Knox}  \&
  {Ma}}{{Santos} et~al.}{2003}]{Santos2003}
{Santos} M.~G.,  {Cooray} A.,  {Haiman} Z.,  {Knox} L.,   {Ma} C.-P.,  2003,
  The Astrophysical Journal

\bibitem[\protect\citeauthoryear{{Simonyan} \& {Zisserman}}{{Simonyan} \&
  {Zisserman}}{2014}]{Simonyan2014}
{Simonyan} K.,  {Zisserman} A.,  2014, arXiv e-prints, p. arXiv:1409.1556

\bibitem[\protect\citeauthoryear{Spitler, Romanowsky, Diemand, Strader, Forbes,
  Moore  \& Brodie}{Spitler et~al.}{2012}]{Spitler2012}
Spitler L.~R.,  Romanowsky A.~J.,  Diemand J.,  Strader J.,  Forbes D.~A.,
  Moore B.,   Brodie J.~P.,  2012, Monthly Notices of the Royal Astronomical
  Society, 423, 2177

\bibitem[\protect\citeauthoryear{Srivastava, Hinton, Krizhevsky, Sutskever  \&
  Salakhutdinov}{Srivastava et~al.}{2014}]{Nitish2014}
Srivastava N.,  Hinton G.,  Krizhevsky A.,  Sutskever I.,   Salakhutdinov R.,
  2014, Journal of Machine Learning Research, 15, 1929

\bibitem[\protect\citeauthoryear{Su, Gu, III  \& Irwin}{Su
  et~al.}{2014}]{Su2014}
Su Y.,  Gu L.,  III R. E.~W.,   Irwin J.,  2014, The Astrophysical Journal,
  786, 152

\bibitem[\protect\citeauthoryear{{Watkins}, {Evans}  \& {An}}{{Watkins}
  et~al.}{2010}]{Watkins2010}
{Watkins} L.~L.,  {Evans} N.~W.,   {An} J.~H.,  2010, Monthly Notices of the
  Royal Astronomical Society, 406, 264

\makeatother
\end{thebibliography}
\bibliographystyle{mnras}

\end{document}